\documentstyle[11pt,aasms4]{article}

\newcommand{\be}{\begin{eqnarray}}
\newcommand{\ee}{\end{eqnarray}}

\newcommand\ie {{\it i.e.}}
\newcommand\eg {{\it e.g.}}
\newcommand\etal{{\it et al.}}

\newcommand\simgreater{\,\lower0.7ex\hbox{$\stackrel{>}{\sim}$}\,}
\newcommand\simless{\,\lower0.7ex\hbox{$\stackrel{<}{\sim}$}\,}

\newcommand{\nue}{$\nu_{{\rm e}}$}
\newcommand{\nuebar}{$\bar{\nu}_{{\rm e}}$}

\begin{document}

\title{
An Investigation of Neutrino-Driven Convection\linebreak and the Core Collapse Supernova Mechanism 
\linebreak Using Multigroup Neutrino Transport
}

\author{A. Mezzacappa\altaffilmark{1,2}, 
A. C. Calder\altaffilmark{1,3}, 
S. W. Bruenn\altaffilmark{4}, 
J. M. Blondin\altaffilmark{5},\linebreak 
M. W. Guidry\altaffilmark{1,2},
M. R. Strayer\altaffilmark{1,2}, and 
A. S. Umar\altaffilmark{3}}
\vspace{3.5in}
\begin{center}
Accepted for publication in {\it The Astrophysical Journal}
\end{center}

\altaffiltext{1}{Theoretical and Computational Physics Group, Oak Ridge National
Laboratory, Oak Ridge, TN 37831--6373}
\altaffiltext{2}{Department of Physics and Astronomy, University of Tennessee,
Knoxville, TN 37996--1200}
\altaffiltext{3}{Department of Physics and Astronomy, Vanderbilt University,
Nashville, TN 37235}
\altaffiltext{4}{Department of Physics, Florida Atlantic University, Boca Raton, FL 33431--0991}
\altaffiltext{5}{Department of Physics, North Carolina State University, 
Raleigh, NC 27695--8202}
\newpage

\begin{center}
{\large {\bf Abstract}}
\end{center}

We investigate neutrino-driven convection in core collapse supernovae and 
its ramifications for the explosion mechanism. We begin with an ``optimistic'' 
postbounce model in two important respects: (1) we begin with a 15 M$_{\odot}$ 
precollapse model, which is representative of the class of stars with compact 
iron cores; (2) we implement Newtonian gravity. Our precollapse model is 
evolved through core collapse and bounce in one dimension using multigroup 
(neutrino-energy--dependent) flux-limited diffusion ({\small MGFLD}) neutrino 
transport and Newtonian Lagrangian hydrodynamics, providing realistic initial 
conditions for the postbounce convection and evolution. 

Our two-dimensional simulation began at 12 ms after bounce and proceeded 
for 500 ms. We couple two-dimensional ({\small PPM}) hydrodynamics to 
precalculated one-dimensional {\small MGFLD} neutrino transport. (The
neutrino distributions used for matter heating and deleptonization 
in our {\small 2D} run are obtained from an accompanying {\small 1D} simulation. The
accuracy of this approximation is assessed.) For the moment we
sacrifice dimensionality for realism in other aspects of our neutrino
transport. {\small MGFLD} is an implementation of neutrino transport 
that simultaneously (a) is multigroup and (b) simulates with sufficient 
realism the transport of neutrinos in opaque, semitransparent, and 
transparent regions. Both are crucial to the accurate determination 
of postshock neutrino heating, which sensitively depends on the luminosities, 
spectra, and flux factors of the electron neutrinos and antineutrinos 
emerging from their respective neutrinospheres. 

By 137 ms after bounce, we see neutrino-driven convection rapidly developing
beneath the shock. By 212 ms after bounce, this convection becomes large-scale, 
characterized by higher-entropy, expanding upflows and denser,
lower-entropy, finger-like downflows. The upflows reach the shock and distort it from
sphericity.  The radial convection velocities at this time become supersonic just below
the  shock, reaching magnitudes in excess of $10^{9}$ cm/sec. Eventually, however, 
the shock recedes to smaller radii, and at $\sim$500 ms after bounce there is 
no evidence in our simulation of an explosion or of a developing explosion. 

Our angle-averaged density, entropy, electron fraction, and radial velocity
profiles in our two-dimensional model agree well with their counterparts
in our accompanying one-dimensional {\small MGFLD} run above and below
the neutrino-driven convection region. In the convective region, the one-dimensional
and angle-averaged profiles differ somewhat because (1) convection tends
to flatten the density, entropy, and electron fraction profiles, and
(2) the shock radius is boosted somewhat by convection. However, 
the differences are not significant, indicating that, while vigorous,
neutrino-driven convection in our model does not have a significant
impact on the overall shock dynamics.

The differences between our results and those of other groups are 
considered. These most likely result from differences in (1) numerical
hydrodynamics methods, (2) initial postbounce models, and most 
important, (3) neutrino transport approximations. 
We have compared our neutrino luminosities, {\small RMS} energies, and inverse 
flux factors with those from the exploding models of other groups. Above all, 
we find that the neutrino {\small RMS} energies computed by our multigroup 
({\small MGFLD}) transport are significantly lower than the values obtained 
by Burrows \etal\ , who specified their neutrino spectra by tying the 
neutrino temperature to the matter temperature at the neutrinosphere 
and by choosing the neutrino degeneracy parameter arbitrarily, and by Herant 
\etal\ in their transport scheme, which (1) is gray and (2) patches together 
optically thick and thin regions. The most dramatic difference between our 
results and those of Janka and M\"{u}ller is exhibited by the 
difference in the net cooling rate below the gain radii: Our rate is 
2--3 times greater during the critical 50--100 milliseconds after bounce.

We have computed the mass and internal energy in the gain region as 
a function of time. Up to $\sim$ 150 ms after bounce, we find that both increase as a
result of the increasing gain region volume, as the gain and shock
radii diverge. However, at all subsequent times we find that the
mass and internal energy in the gain region decrease with time in
accordance with the density falloff in the preshock region and with
the flow of matter into the gain region at the shock and out of the
gain region at the gain radius. Therefore,
we see no evidence in the simulations presented here that neutrino-driven convection leads to
mass and energy accumulation in the gain region.

We have compared our {\small 1-} and {\small 2D} densities, temperatures, and electron fractions 
in the region below the electron neutrino and antineutrino gain radii, above which 
the neutrino luminosities are essentially constant (i.e., the neutrino sources are 
entirely enclosed), in an effort to assess how spherically symmetric our neutrino 
sources remain during our {\small 2D} evolution, and therefore, to assess our use of 
precalculated {\small 1D MGFLD} neutrino distributions in calculating the matter
heating and deleptonization. We find no differences below the neutrinosphere 
radii, and between the neutrinosphere and gain radii, no differences with obvious ramifications 
for the supernova outcome. We note that the interplay between
neutrino transport and convection below the neutrinospheres is a delicate matter, 
and is discussed at greater length in another paper (Mezzacappa \etal\ 1997a). 
However, the results presented therein do support our use of precalculated 
{\small 1D MGFLD} in the present context.
 
Failure in our ``optimistic'' 15 M$_{\odot}$ Newtonian model leads us to
conclude that it is unlikely, at least in our approximation, that neutrino-driven 
convection will lead to explosions for more massive stars with fatter iron 
cores or in cases in which general relativity is included. 


\newpage
\section{\bf Introduction}

\subsection{Convection and the Supernova Mechanism}

Much of the current research on the core collapse supernova mechanism is focused on the role of
convection. This is motivated in part by a number of observations of SN 1987A, which indicate
that extensive mixing occurred throughout much of the ejected material, which, by inference,
points to fluid instabilities arising during the explosion itself. The early
(July, 1987) detection of continuum X-rays and $\gamma$-rays (Dotani \etal\ 1987; Sunyaev \etal\
1987) and of the 847 and 1238 keV $^{56}$Co $\gamma$-ray lines (August, 1987) (Matz, Share, \&
Chupp 1988; Matz \etal\ 1988), confirmed shortly thereafter by others 
(
Cook \etal\ 1988; 
Gehrels, Leventhal, \& MacCallum 1988; 
Mahoney \etal\ 1988;
Sandie \etal\ 1988; 
Wilson \etal\ 1988;
Teegarden \etal\ 1989
), suggests that mixing of the $^{56}$Co in the ejecta must have occurred.
$^{56}$Co arises from $^{56}$Ni, which is created in the deep silicon layer. Without
the breaking of spherical symmetry, $^{56}$Co and other radioactive elements would remain buried
under the massive stellar envelope for about a year, until the latter became  transparent by expansion
(
Gehrels, MacCallum, \& Leventhal 1987; 
McCray, Shull, \& Sutherland 1988; 
Pinto \& Woosley 1988;
Xu \etal\ 1988 
). Even more direct evidence of significant mixing is the very high Fe velocities
(\simgreater 3000 km s$^{-1}$) inferred from infrared observations of $^{56}$Fe II lines 
(
Erikson \etal\ 1988; 
Rank \etal\ 1988; 
Hass \etal\ 1990;
Spyromilio, Meikle, \& Allen 1990 
), the very
large $^{56}$Co velocities inferred from the 847 keV line profile by the GRIS experiment
(Tueller \etal\ 1990), and hydrogen velocities as low as 800 km s$^{-1}$ (H\"{o}flich 1988). These
observables are evidence that some of the $^{56}$Ni (the progenitor of $^{56}$Co and $^{56}$Fe)
was mixed out to the hydrogen envelope. Because this degree of mixing
is not reproduced by simulations in which the mixing arises only from Rayleigh--Taylor instabilities
in the expanding envelope 
(
Arnett, Fryxell, \& M\"{u}ller 1989; 
Den, Yoshida, \& Yamada 1990; 
Hachisu \etal\ 1990; 
Yamada, Nakamura, \& Oohara 1990; 
Fryxell, Arnett, \& M\"{u}ller 1991; 
M\"{u}ller, Fryxell, \& Arnett 1991; 
Herant \& Benz 1991, 1992
), it must 
be presumed that these instabilities were preceded by a prior round of instabilities, most likely
occurring in the explosion itself (Herant \& Benz 1992).

In addition to the observational evidence of extensive mixing due to fluid instabilities, there are 
compelling theoretical reasons to consider convection. The first and obvious reason is that there
are several unstable regions that develop in the postcollapse core, both below and above the
neutrinosphere (Epstein 1979; Arnett 1986, 1987; Bethe, Brown, \& Cooperstein 1987; Burrows 1987;
Bethe 1990). The second is the failure to explode of supernova simulations that do not incorporate fluid
instabilities (Bruenn 1993; Cooperstein 1993; Wilson and Mayle 1993), and the expectation
that fluid instabilities, by enhancing the transport of lepton-rich matter to the neutrinosphere or
augmenting the neutrino energy deposition efficiency above the neutrinosphere, will be helpful
in generating explosions. These expectations derive from the nature of the explosion ``power up'' phase,
as envisioned by the current core collapse supernova paradigm. According to this paradigm, which is
referred to as the ``shock reheating mechanism'' or ``delayed mechanism'' (Wilson 1985; Bethe and
Wilson 1985), the shock launched into the outer core at core bounce stalls between 100 and 200
km due to nuclear dissociation and neutrino radiation. The shock then becomes an accretion shock,
and within 10's of milliseconds, a quasi--steady-state structure is established in which infalling matter
encountering the shock is dissociated into free nucleons and then heated by the transfer and
deposition of energy by neutrinos radiating from the hot contracting core. As this matter
continues to flow inward, neutrino and compressional heating increase its temperature until the
cooling rate, which goes as the sixth power of the temperature, exceeds the heating rate. The
inflowing matter thereafter cools and eventually accretes onto the core. The radius at
which the heating and cooling rates are equal is referred to as the ``gain radius.'' If the neutrino
heating is sufficiently rapid in the region between the shock and the gain radius, the increased
thermal pressure behind the shock will allow it to overcome the accretion ram pressure and propagate out
through the envelope, thus producing a supernova.

\subsection{Fluid Instabilities: Preliminaries}

We emphasize again that spherically symmetric supernova simulations that have not incorporated fluid
instabilities fail to provide the necessary heating to revive the shock.
To appreciate the potential role of fluid instabilities 
in reviving the shock, we begin with the fact that the energy transfer between neutrinos and matter 
behind the shock is mediated primarily by the charged current reactions:

\begin{equation}
\nu_{\rm e} + {\rm n} \rightleftharpoons {\rm p} +{\rm e^{-}}, \hspace{0.5in}
\bar{\nu}_{\rm e} + {\rm p} \rightleftharpoons {\rm n} + {\rm e^{+}}.
\label{eq:r1}
\end{equation}

\noindent It follows that the heating and cooling rates by \nue's are given by

\begin{equation}
\frac{ {\rm Heating} }{ {\rm Nucleon} } \propto  L_{\nu_{\rm e}}
\langle \epsilon_{\nu_{\rm e}}^{2} \rangle
\langle \frac{1}{ {\cal F}_{\nu_{{\rm e}}} } \rangle
\label{eq:e1}
\end{equation}

\begin{equation}
\frac{ {\rm Cooling} }{ {\rm Nucleon} } \propto T_{\rm m}^{6}
\label{eq:e2}
\end{equation}

\noindent with similar expressions for the \nuebar's. Here, $L_{\nu_{\rm e}}$ is the \nue\
luminosity; $\langle \epsilon_{\nu_{\rm e}}^{2} \rangle$ is the \nue\ mean square energy, averaged
with respect to $\epsilon_{\nu_{\rm e}}^{3}$; $\frac{1}{ {\cal F}_{\nu_{{\rm e}}} } \simgreater 1$ is
the inverse flux factor, which equals $c \times U_{\nu_{\rm e}}/F_{\nu_{\rm e}}$, where
$U_{\nu_{\rm e}}$ and $F_{\nu_{\rm e}}$ are the \nue\ energy density and flux, respectively; and
$T_{\rm m}$ is the local matter temperature. We have neglected the electron degeneracy of the
matter in equation (\ref{eq:e2}). 

\subsection{Fluid Instabilities below the Neutrinosphere}

Fluid instabilities occur in several distinct regions in a postcollapse stellar core. Near and below
the neutrinosphere, the dissipation of the shock due to nuclear dissociation and \nue\ radiation will
imprint a negative entropy gradient, and therefore destabilize this region to entropy-driven convection
(Arnett 1986, 1987; Burrows 1987). While the convection will not be sustained, it can lead to a rapid
initial turnover of the region. The material near and below the neutrinosphere will also be
destabilized by a negative lepton gradient (Epstein 1979). This results from the lepton ``trough''
produced near the neutrinosphere by the combination of rapid electron capture and \nue\ escape. A
negative lepton gradient connects this region with the more lepton-rich material at smaller radii.
Because electron capture and \nue\ radiation continues near the neutrinosphere 
as the protoneutron star evolves, the tendency
of lepton-driven fluid motions to flatten the negative lepton gradient will be resisted, and the
instability driving these fluid motions may tend to persist. Deeper in the core, the destabilizing
effect of the negative lepton gradient will be counteracted by the stabilizing effect of the positive
entropy gradient left over from the shock as it propagates through this region while gathering
strength from the rebounding inner core. With the diffusion of both energy and leptons by
neutrinos, this region can be destabilized on a diffusion timescale, if the diffusion rates for energy
and leptons are different. Mayle (1995) and Wilson and Mayle (1988, 1993) have argued that the material
in this region will be unstable to one of these doubly diffusive instabilities referred to as ``neutron
fingers.'' Bruenn, Mezzacappa and Dineva (1995) and Bruenn and Dineva (1996) have disputed this claim.
It is not our purpose here to delve deeper into the nature or existence of the possible modes of
fluid instabilities below the neutrinosphere. Let us simply refer in the following to
the fluid motions that would result from any of the above fluid instabilities as ``protoneutron star
convection.'' In most cases, protoneutron star convection will be confined to the region below and
encompassing the neutrinosphere.

The effect of protoneutron star convection on shock heating can be ascertained by referring to
equation (\ref{eq:e1}), where it is seen that the heating rate depends on the product of the
factors $L_{\nu_{\rm e}}$, $\langle \epsilon_{\nu_{\rm e}}^{2} \rangle$, and $\langle \frac{1}
{ {\cal F}_{\nu_{{\rm e}}} } \rangle$. The last of these factors, the inverse flux factor, is primarily
geometrical, depending on the ratio of the neutrinosphere radius to the radial distance of the
fluid element in question. It approaches unity, as this ratio becomes small, reflecting the fact that
the neutrinos are radially free streaming in this limit. However, the first two factors depend on
the conditions at the neutrinosphere, and can therefore be affected by protoneutron star convection.
Let us suppose that the neutrinosphere lies near the bottom of a negative entropy gradient. Then 
entropy-driven protoneutron star convection will advect high-entropy material from deeper regions, up to the
vicinity of the neutrinosphere, raising the temperature there. This will clearly increase both
$L_{\nu_{\rm e}}$ and $\langle \epsilon_{\nu_{\rm e}}^{2} \rangle$, with similar consequences for
the corresponding \nuebar\ quantities. (This is obvious in the limit of zero electron degeneracy,
where the product of these two factors is proportional to $T^{6}_{\nu_{{\rm e}}}$, where
$T_{\nu_{{\rm e}}}$ is the \nue-sphere temperature.) In addition to increasing the neutrinosphere
temperature, protoneutron star convection can also increase the neutrinospheric lepton fraction. \
This follows from the fact that rapid electron capture and \nue\ escape at the neutrinosphere
causes it to lie near a lepton fraction ($Y_{\ell}$) minimum. Entropy- or lepton-driven protoneutron
star convection will advect lepton-rich matter to the vicinity of the neutrinosphere, and thereby
increase the electron degeneracy there. This will increase both $L_{\nu_{\rm e}}$ and $\langle
\epsilon_{\nu_{\rm e}}^{2} \rangle$, while decreasing $L_{\bar{\nu}_{\rm e}}$ and $\langle
\epsilon_{\bar{\nu}_{\rm e}}^{2} \rangle$. In most cases, the net effect will be to augment the
heating rate of the material behind the shock.

Since the original suggestion of Epstein (1979), protoneutron star convection has been studied by a
number of groups 
(
Arnett 1986, 1987; 
Bethe 1990, 1993; 
Bruenn and Mezzacappa 1994; 
Bruenn, Mezzacappa, and Dineva 1995; 
Bruenn and Dineva 1996; 
Burrows 1987; 
Burrows and Lattimer 1988; 
Burrows and Fryxell 1992, 1993; 
Burrows, Hayes, and Fryxell 1995; 
Colgate, Herant, and Benz 1993; 
Herant, Benz, and Colgate 1992; 
Herant \etal\ 1994; 
Janka and M\"{u}ller 1993, 1995, 1996; 
Keil, Janka, and M\"{u}ller 1996;
Mezzacappa \etal\ 1997a; 
M\"{u}ller 1993; 
M\"{u}ller and Janka 1994; 
Wilson and Mayle 1988, 1993; 
). 
Despite this work, the nature and extent of protoneutron star convection and its
effect on shock heating is still unclear. The problem is that the coupling between neutrino transport
and the fluid motions arising from instabilities in this region is strong, requiring that realistic
simulations incorporate both multi-D hydrodynamics and multi-D multigroup neutrino transport. Until this has
been accomplished, the picture will probably remain unclear.

\subsection{Fluid Instabilities above the Neutrinosphere}

As infalling material encounters the shock, it is dissociated into free neutrons and protons
if the shock is within a radius $\sim 200$ km. (At larger radii there will be an admixture of
alpha particles.) As this material continues to flow inward, it will be heated by the charged-current
reactions [equation (\ref{eq:r1})], until it reaches the gain radius. Furthermore, the neutrino
heating is strongest just beyond the gain radius and decreases farther out, as the neutrino flow
becomes radially diluted. These two factors conspire to create a negative entropy
gradient between the gain radius and the shock, which will be unstable to entropy-driven 
convection. Because this convection is driven by neutrino heating, it will persist as long
as neutrinos heat, or until an explosion develops. This convection, which is the subject of
this paper, will be referred to as ``neutrino-driven convection'' or ``{\small ND} convection.''

Bethe (1990) pointed out that {\small ND} convection would reduce or eliminate the density inversions
found at the outer edge of the hot bubble that formed in the successful supernova simulations
of Wilson and Mayle (1993). {\small ND} convection was not incorporated in these calculations, and an enormous
negative density gradient developed behind the shock as it propagated out. He noted also that
neutrino-driven convection would bring cooler material down to the vicinity of the gain 
radius where it would be more efficiently heated, because the cooling rate, as given by equation
(\ref{eq:e2}), would be reduced; the outward flow of hot material would bring energy to the shock,
and thus help to support it.

The first {\small 2D} supernova simulation incorporating {\small ND} convection was performed by Herant, Benz,
and Colgate (1992) using a smooth-particle--hydrodynamics ({\small SPH}) code. 
They found that, if {\small ND} convection
is able to develop, it plays a crucial role in generating an efficient explosion. The convective
flows in their calculations became large-scale ``long-wavelength'' flows, providing 
an efficient (separate hot and cold matter) way of conveying low-entropy matter from the shock to the gain
radius and high-entropy matter back to the shock. Furthermore, the latent heat of the 
alpha--free-nucleon transition enables the storage and transport of large amounts of energy without
large temperature increases. They noted finally that {\small ND} convection would allow accretion to
continue while the shock is moving outward in radius, thereby maintaining an energy input
to the shock from the neutrinos radiated by the accreting matter. Colgate, Herant, and Benz (1993)
examined the accretion onto the neutron star during the explosion [the Herant, Benz, and
Colgate (1992) simulations could not resolve this aspect of the problem, and they ignored the
contribution to the neutrino luminosity by the accreting matter], and suggested that accretion would
be unstable, leading to episodic bursts of high-energy neutrinos that would be efficiently absorbed
by the overlying material. This conjecture has not yet been confirmed. In a set of
more refined and self-consistent calculations, Herant \etal\ (1994) confirmed the conclusions
of Herant, Benz, and Colgate (1992), and emphasized the large-scale character of {\small ND} convection
and its role in producing robust and self-regulated supernova explosions. They concluded
that {\small ND} convection is a powerful ``convective engine'' that feeds energy into the shock until
the required explosion energy has been reached. Thus, {\small ND} convection almost guarantees a successful
explosion.

A simulation using entirely different numerical techniques by Miller, Wilson, and Mayle (1993)
reached a very different conclusion. They used a semi-Eulerian finite-difference method (Bowers
and Wilson 1991), and a combination of light-bulb and two-temperature diffusion schemes for neutrino
transport, calibrated by their sophisticated {\small 1D} code. As in the Herant, Benz, and Colgate (1992)
and the Herant \etal\ (1994) simulations, they found that {\small ND} convection tended to evolve to a
large-scale, long-wavelength flow. However, the growth rates for their convection were much slower, 
\eg, an e-folding time of 13 ms, versus a fully developed convection in 20--25 ms after bounce. 
Because their convection
growth rates were slow, and because the unstable conditions for convection's growth are not
well established until $\sim 0.1$ second after bounce, Miller, Wilson, and Mayle (1993) found
that {\small ND} convection was not particularly important in their simulations. Acknowledging that the
limitations of their {\small 2D} model favored explosions, they concluded that more realistic
{\small 2D} simulations would show that the effects of {\small ND} convection are unlikely to revive a stalled
shock.

{\small 2D} supernova simulations using hydrodynamics codes based on the piecewise-parabolic--method 
({\small PPM})
of Colella and Woodward (1994) and incorporating {\small ND} convection were performed by Burrows, Hayes, 
and Fryxell (1995), and by Janka and M\"{u}ller (1995, 1996). In contrast to the results of Herant 
\etal\ and Miller \etal, the {\small PPM} calculations found {\small ND} convection to be more turbulent, 
although a large-scale flow was exhibited. The flows consisted of 
high-entropy rising bubbles and balloons, distorted into mushrooms by Kelvin-Helmholtz instabilities, 
and low-entropy descending narrow flux tubes; and the mode numbers ranged from 2 to 10. Burrows, Hayes,
and Fryxell (1995) implemented radial transport along their angular rays in a gray diffusion
approximation for optical depths $\geq$ 1/2, with a neutrino-matter coupling for optical depths 
$\leq$ 1/2 that assumes particular electron neutrino and antineutrino energy spectra. 
Unlike the Herant \etal\ simulations, they found that neither mass nor
energy accumulate in the convection region. They were unable to verify the basic ``convective
engine'' paradigm of Herant \etal\ and found, instead, that if an explosion ensues, it is
because of the decline in the accretion ram, rather than an increase in the shock energy
[see also Burrows and Goshy (1993)].
They found that {\small ND} convection allows higher entropies to develop in the convective region
than in {\small 1D} simulations because of the longer time a given fluid element spends in the gain
region (1 or 2 cycle times before passing inward through the gain radius). This combined
with the dynamic pressure of the buoyant plumes causes the stalled shock radius to
equilibrate at a larger value and therefore at a lower gravitational potential. The
corresponding reduction in the accretion ram makes it easier for {\small 2D} supernova models to
explode, but does not guarantee an explosion. Reducing the magnitudes of the neutrino-matter
coupling terms by assuming a different neutrino spectrum led to failures.

Janka and M\"{u}ller (1995, 1996) performed a very useful parameter study of {\small 1D} and {\small 2D}
supernova simulations by varying a light-bulb neutrino source. Like Burrows \etal\ , and unlike
Herant \etal\ , they did not find an accumulation of energy in the {\small ND} convection region until
the typical value of 10$^{51}$ erg was reached. However, unlike the Burrows \etal\ simulations, 
their simulations did not exhibit the vigorous boiling that immediately preceded an explosion.
They also found that accretion onto the protoneutron star ceased when an explosion got
underway; therefore, the explosion was not continuously fed by accretion luminosity. They did
find that {\small ND} convection leads to a higher net efficiency of neutrino energy deposition and
an efficient mechanism of transporting energy to the shock. Their overall conclusions were
that {\small ND} convection is important for generating explosions only in a rather narrow window
($\Delta L/L \sim 20$ \%) of neutrino luminosities. Below this window, neither {\small 1D} 
nor {\small 2D} simulations
would explode; above this window, both would explode.

\subsection{This Work}

It is apparent from the above discussion that there is considerable disagreement regarding the role
of {\small ND} convection in the supernova mechanism. All of the above simulations were pioneering
in the sense that multi-D hydrodynamics was used. However, computer limitations necessitated
that rather severe approximations be made in the neutrino transport and the coupling of
neutrinos to matter. We feel that much of the disparity in the above results can be traced
to differences in the treatment of neutrinos. In fact, we will show in this paper that
the supernova simulations incorporating {\small ND} convection that give rise to explosions
assume a hard neutrino spectrum (thus favoring explosions) when compared with multigroup 
calculations.

The purpose of this paper is to eliminate some of the uncertainties associated with neutrino
transport approximations by ``coupling'' (the meaning of this term will be made clear below)
a sophisticated multigroup flux-limited diffusion ({\small 
MGFLD}) neutrino transport code with the numerically nondiffusive high-order {\small PPM} hydrodynamics
code, EVH-1, to examine the role of {\small ND} convection in the supernova mechanism. The importance
of using multigroup neutrino transport is that the neutrino energy spectrum is part of the
solution and need not be assumed. The importance of using flux-limited diffusion is that the
transport of neutrinos from optically thick to optically thin regions is computed seamlessly
and with sufficient realism. 

Our procedure, described in more detail in Mezzacappa \etal\ (1997a), is to perform a 
{\small 1D} (Lagrangian) simulation of core collapse and the postbounce evolution to (1) generate running
(\ie\ , time-dependent) inner and outer boundary conditions for all of the relevant variables
at our fixed inner and outer Eulerian boundaries, and (2) generate tables of
the zeroth angular moments of the neutrino distributions as a function of time, radius,
and neutrino energy. The {\small 2D} simulation is then carried out using the inner and outer
boundary conditions generated by our {\small 1D} simulation, and the neutrino moment tables are
used to compute the local energy and lepton exchange between the matter and the neutrinos.

Because {\small ND} convection occurs between the gain radius and the shock, the convecting material has 
small neutrino optical depths for all of the relevant neutrino energies. Thus, we expect the 
feedback between the hydrodynamics and the neutrino transport in this region to be minimal. 
If this were the entire story, coupling {\small 1D MGFLD} to {\small 2D} hydrodynamics would be 
an excellent approximation. However, the neutrino radiation field in the gain region is
largely determined by the neutrino transport below it, particularly above and below the
neutrinospheres. Therefore, to fully assess the accuracy of {\small 1D} transport, we must consider 
the feedback on the radiation field that results from (1) asymmetric accretion through the 
gain radius from the neutrino-driven convection region and (2) nonspherically symmetric 
structure below the neutrinospheres that may result, for example, from protoneutron star
convection [see Wilson and Mayle (1993), Herant \etal\ (1994), Burrows \etal\ (1995), Keil 
\etal\ (1996), and Mezzacappa \etal\ (1997a)]. We consider the first effect here. The second
effect has been considered at length in Mezzacappa \etal\ (1997a), and the key results 
relevant to the claims made in this paper are mentioned again here (in Section 4). 

In Section 2, we briefly describe our initial models, codes, and methodology.
Our results are then presented in Section 3. Section 4 is devoted to an assessment 
of our {\small 1D MGFLD} neutrino transport approximation, and in Section 5 we
summarize our results, compare them with other groups, and state our conclusions. 

\newpage
\section{Initial Models, Codes, and Methodology}

We begin with the 15 ${\rm M}_{\odot}$ precollapse model S15s7b 
(Woosley and Weaver 1995, Weaver and Woosley 1997). The initial model was evolved through 
core collapse and bounce using {\small MGFLD} neutrino transport 
and Lagrangian hydrodynamics, providing realistic initial conditions
for the postbounce convection and evolution. The one-dimensional 
data at 12 ms after bounce (211 ms after the initiation of core 
collapse) were mapped onto our two-dimensional Eulerian grid. 
The inner and outer boundaries of our grid were chosen to be at 
radii of 20 km and 1000 km, respectively. 128 nonuniform radial 
spatial zones were used, and 128 uniform angular zones spanning 
a range of 180 degrees were used for $\theta$ (together with 
reflecting boundary conditions). 

Because the finite differencing in our {\small PPM} scheme is
nearly noise free, and because we cannot rely on machine roundoff
to seed convection in a time that is short compared with the 
hydrodynamics time scales in our runs, we seeded convection 
everywhere on the grid by applying random velocity perturbations
to the radial and angular velocities, between  $\pm$5\% of the
local sound speed. Our seeding included the initial Ledoux 
unstable regions below and around the neutrinospheres immediately 
following core bounce.

Time-dependent inner and outer boundary data for the enclosed mass, density,
temperature, electron fraction, pressure, specific internal energy,
and velocity were supplied by our accompanying one-dimensional 
{\small MGFLD} run, which was continued for this purpose for 700 ms
after bounce. The inner and outer boundaries
were chosen to be at 20 km (deep within the core) and at 1000 km 
(well outside the shock), respectively, which are regions in which the flow is
spherically symmetric. 

It is also important to note here that for the matter heating and
deleptonization in our {\small 2D} run, we use tables of precalculated 
neutrino distributions, $\psi^{0}_{\nu_{\rm e},\bar{\nu}_{\rm e}}
(r,t;E_{\nu})$, obtained from our accompanying {\small 1D} run that implements 
{\small MGFLD} and Lagrangian hydrodynamics. Both our {\small 1-} and {\small 2D}
simulations are Newtonian. The accuracy of using precalculated 
{\small 1D} neutrino distributions in our {\small 2D} models is discussed in Section 
4. 

Details of the codes used in our simulations, and more detail 
on our methodology, can be found in Mezzacappa \etal\ (1997a).

\newpage
\section{\bf Results}

Figure 1 shows the velocity, density, entropy, and electron fraction
profiles at the start of our run, 12 ms after bounce. 

In Figures 2a--d we plot the results of one-dimensional simulations
using our {\small MGFLD} code and our {\small PPM} hydrodynamics code. 
(The PPM code has been generalized for 
realistic equations of state and for neutrino heating and  cooling and 
deleptonization). We compare density, entropy, electron fraction, and 
velocity at two different times during the course of
a 500 ms run. The figures illustrate that the agreement is excellent,
with the primary differences resulting from our {\small PPM} code's
ability to better resolve the shock. This gives us confidence that
(1) the neutrino heating and cooling and deleptonization are 
simulated well by our {\small PPM} code, and (2) that differences 
between two-dimensional and one-dimensional simulations of the shock 
dynamics will result from convection's presence in the former case, 
not from numerical sources. 

Figure 3a shows the two-dimensional entropy profiles at three select
times during our 500 ms run. At 137 ms after bounce, neutrino-driven
convection is rapidly developing below the shock, having entered the
nonlinear regime. There is evidence of expanding higher-entropy rising 
flows, and denser lower-entropy infalling matter. Roughly 7 rising 
plumes can be counted at this time. The convection has not yet reached 
the shock. By 212 ms 
after bounce, neutrino-driven convection is fully developed. A clear 
contrast is evident between the higher-entropy upflows and the lower-entropy 
downflows. At this time, the rising plumes have reached the shock and 
distorted it, and the convection is semiturbulent, although only 4 
(versus 7) plumes in a range of 180 degrees can be counted now. Although 
more turbulent, our flow patterns are not in complete disagreement 
with the flows obtained by another group using {\small SPH} (Herant \etal\ 
1994); the latter obtain a more orderly low-mode convection. However, 
differences between the {\small SPH} and {\small PPM} simulations are 
significant and most likely have important dynamical consequences. The low-mode 
convection obtained by Herant \etal\ led them to develop their ``thermodynamic 
engine'' interpretation of the supernova mechanism, although this ``engine'' 
apparently did not work in the Miller \etal (1993) simulations, 
in which low-mode convection was also exhibited. Such an interpretation
is even less possible in the {\small PPM} simulations, where high- and low-entropy 
flows are not as well separated. We also mention here that our flow 
patterns are in qualitative agreement with those obtained by other 
groups that implement {\small PPM} hydrodynamics (Burrows \etal\ 1995, 
Janka and M\"{u}ller 1996). Finally, at 512 ms after bounce, Figure
3a illustrates that our shock has receded to smaller radii, and that
the convection beneath it has become even more turbulent, mixing
to an even greater extent high- and low-entropy matter. It is clear 
that, even in the presence of large-scale neutrino-driven convection,
we do not, with our {\small MGFLD} neutrino transport, obtain an 
explosion. 

In Figure 3b, we plot the one-dimensional entropy and two-dimensional angle-averaged entropy, 
both computed by the {\small PPM} code, 
as a function of radius at the same three time slices. 
The angle-averaged entropy is defined by 
\smallskip

\begin{equation}
< S >(i) = \frac{1}{A(i)} \sum^{n_{\theta}}_{j = 1} A(i,j) S(i,j)
\label{eq:avgent}
\end{equation}
\smallskip

\noindent where
\smallskip

\begin{equation}
A(i,j)=2\pi r^{2}(i)\sin\theta(j)d\theta
\label{eq:area}
\end{equation}
\smallskip

\noindent and where $A(i)=4\pi r^{2}(i)$ and $d\theta =\pi/128$.
At 137 ms after bounce, as expected, there is very little difference between the 
one- and two-dimensional profiles --- convection has not yet fully developed and
one would not expect large differences between the two simulations. At 212 ms after 
bounce, when convection has fully developed, it has clearly flattened the peak in 
the entropy profile between 90 and 150 km, relative to the one-dimensional case, and it 
is also clear that the shock location is a bit farther out --- convection does seem to 
have an effect on the shock radius in our simulations, albeit marginal, pushing it out 
to slightly larger radii. Moreover, at this time, the maximum angle-averaged entropy is 13, 
which is not particularly high, nor even marginal for an explosion. For example, Burrows \etal\ (1995) 
find that the entropies in their rising bubbles reach 25--35 prior to their  
explosions. Under explosive conditions, we expect entropies between the cooling 
protoneutron star and the shock to rise well past $\sim$ 30, a condition signaling the 
formation of a low-density radiation-dominated ``bubble''  behind the shock, as the shock 
separates itself from the protoneutron star. At the last time slice, 512 ms after 
bounce, the shock front is located at smaller radii, and the entropy jump across 
it is greater --- the entropy behind the shock rises in time because the preshock
matter density decreases and because the gravitational potential well becomes
deeper, as more material accumulates onto the nascent neutron star and 
the radius of the shock decreases. 
The peak entropies for this last slice reach 17--18. In addition,
the differences between our one- and two-dimensional simulations are less pronounced
than at $t_{\rm pb}=212$ ms. 

In Figure 4a, we show the two-dimensional electron fraction profiles at the same
three postbounce times. These profiles are complementary to the entropy profiles.
In particular, at $t_{\rm pb}=512$ ms we see low-$Y_{\rm e}$ (high-entropy) matter 
rising in dramatic expanding plumes, while high-$Y_{\rm e}$ (low-entropy) matter 
infalls in dense finger-like flows. The $Y_{\rm e}$ contrast between rising and 
falling flows is greatest at $t_{\rm pb}=137$ ms, when convection is developing 
and matter at disparate $Y_{\rm e}$ values in the $Y_{\rm e}$ profile beneath 
the shock, extending down to the trough near the neutrinospheres, is affected. 
In time, however, this contrast is lessened, and finally at $t_{\rm pb}=512$ ms, 
the matter in the neutrino-driven convection region has been sufficiently mixed 
that very little contrast is exhibited. 

Figure 4b plots the angle-averaged electron fraction in our two-dimensional simulation 
against the corresponding one-dimensional results, both obtained with our {\small PPM}
code. This quantity is defined by

\begin{equation}
< Y_{\rm e} >(i) = \frac{1}{A(i)} \sum^{n_{\theta}}_{j = 1} A(i,j) Y_{\rm e}(i,j)
\label{eq:avgye}
\end{equation}
\smallskip

\noindent No differences are seen near the $Y_{\rm e}$ 
trough, beneath the neutrino-driven convection region; all differences occur 
directly below the shock, as expected. For example, a flattening in $Y_{\rm e}$ 
is evident between $t_{\rm pb}=137$ ms and $t_{\rm pb}=512$ ms, which is clearly seen 
in the region between 90 and 150 km at $t_{\rm pb}=212$ ms. In addition, in 
both one and two dimensions, the electron fraction gradient steepens as the shock 
recedes and high-$Y_{\rm e}$ matter advects inward through it. Similar to what we 
found in comparing the one- and two-dimensional entropy profiles, very little 
difference between our one- and two-dimensional electron fraction profiles is 
seen at the end of our simulation at $t_{\rm pb}=512$ ms. 

In Figure 5a, we show the two-dimensional radial velocity profiles at $t=$
137, 212, and 512 ms after bounce. Comparing Figure 3a and 5a, high-velocity 
portions of the flow, accelerated by the bouyancy force, are associated both 
with inflow and with outflow. This is most easily seen looking at the second 
panel in both figures. The associations between the high inward radial velocity 
of the finger-like flow at $\theta\approx 60$ degrees and the high-velocity 
outflow associated with a portion of the rising plume at $\theta\approx 120$ 
degrees can easily be made. In Figure 5b, we plot the angle-averaged radial
velocity from our two-dimensional run against the radial 
velocity from our {\small 1D PPM} run, as a function of time over our 500 
ms window. The angle-averaged radial velocity is defined by 
\smallskip

\begin{equation}
< v > \; = \; \frac{1}{A(i)} \sum^{n_{\theta}}_{j = 1} A(i,j) \mid v_r(i,j) \mid
\label{eq:avgvel}
\end{equation}
\smallskip

\noindent The two profiles agree quite well before convection has fully
developed at $t_{\rm pb}=137$ ms; differ most at $t_{\rm pb}=212$ ms, when
convection is fully developed; and are in closer agreement at the end of
our run at $t_{\rm pb}=512$ ms. When convection has had a chance to develop,
the angle-averaged shock radius is farther out relative to the one-dimensional
shock radius, but not significantly so. Moreover, in both simulations, the
shock recedes and strengthens, as evidenced by the increased velocity jump
across it at the end of our run. 
[In both our one- and two-dimensional simulations, there is a small periodic 
inward and outward movement of the shock, \ie\ , a recession and a strengthening, 
followed by an advance and a weakening, and so on. As the shock moves out after 
strengthening, less material is advected onto the protoneutron star, and the 
accretion luminosity drops. This decreases the neutrino heating behind the shock, 
thus undermining its pressure support, and the shock moves inwards. As the shock 
moves inwards, the accretion luminosity rises and the shock is once again strengthened, 
and moves out. This behavior has also been noted by others (\eg\ , Mayle 1985).] 

In Figure 6, we plot the angle-averaged radial and angular convection velocities 
in our two-dimensional run for the three postbounce slices we have focused on in
this discussion, along with the angle-averaged sound speed. The convection velocities 
are defined by 

\begin{equation}
< v_{c} >_{r} = \frac{1}{A(i)} \sum^{n_{\theta}}_{j = 1} A(i,j) \mid \: \mid v_r(i,j) \mid - < v > \: \mid
\label{eq:avgrvel}
\end{equation}
\smallskip

\noindent and
\smallskip

\begin{equation}
< v_{c} >_{\theta} = \frac{1}{A(i)} \sum^{n_{\theta}}_{j = 1} A(i,j) \mid v_\theta(i,j) \mid
\label{eq:avgthetvel}
\end{equation}
\smallskip

\noindent At the time when convection is fully developed, at $t_{\rm pb}=212$ ms,
the radial convection velocities in our simulation become supersonic just below 
the shock. Notice, also, the anticorrelation between the radial and angular convection 
velocity profiles, which is consistent with the convective flow shown, for example,
in Figure 3a. The radial convection velocity is at a minimum at the top of the 
convecting region where the matter turns over, at which point the angular convection
velocity is at a maximum. At $t_{\rm pb}=512$ ms, the radial convection velocity 
is still supersonic just below the shock, despite the recession of the shock
and the more turbulent convection. 

For matter in nuclear statistical equilibrium, the thermodynamic state is
completely determined once the density, temperature (or equivalently, entropy),
and electron fraction are known. In Figures 3b and 4b, we have already graphed 
the angle-averaged entropy and electron fraction from our two-dimensional simulation, 
together with the corresponding profiles from our one-dimensional {\small MGFLD} 
run. For completeness, in Figure 7 we plot the angle-averaged density. In the 
two-dimensional case, at $t_{\rm pb}=212$ and 512 ms, the shock is farther out 
in radius and the density jump therefore occurs at larger radii. Additionally, the
shock is distorted in our two-dimensional simulation, and this gives the appearance 
in the angle-averaged density of shock smearing. Away from the
shock, however, the one- and two-dimensional profiles are very nearly the same. 

The neutrino heating rate (in MeV/nucleon) in the region between the neutrinospheres 
and the shock can be written as [see also equation (2)]

\begin{equation}
\dot{\epsilon}=\frac{X_{n}}{\lambda_{0}^{a}}\frac{L_{\nu_{\rm e}}}{4\pi r^{2}}
                \langle E^{2}_{\nu_{\rm e}}\rangle\langle\frac{1}{\cal F}\rangle
              +\frac{X_{p}}{\bar{\lambda}_{0}^{a}}\frac{L_{\bar{\nu}_{\rm e}}}{4\pi r^{2}}
                \langle E^{2}_{\bar{\nu}_{\rm e}}\rangle\langle\frac{1}{\bar{\cal F}}\rangle
\label{eq:heatrate}
\end{equation}
\smallskip

\noindent where $X_{n,p}$ are the neutron and proton fractions; $\lambda_{0}^{a},
\bar{\lambda}_{0}^{a}$ are the coefficients of the $E_{\nu_{\rm e},\bar{\nu}_{\rm e}}^{-2}$ 
neutrino-energy dependences in the electron neutrino and antineutrino mean free paths, respectively; 
$L_{\nu_{\rm e},\bar{\nu}_{\rm e}}$, $\langle E^{2}_{\nu_{\rm e},\bar{\nu}_{\rm e}}\rangle$, 
and $\langle 1/{\cal F},\bar{\cal F}\rangle$ are the electron neutrino and antineutrino luminosities, 
mean square energies, and mean inverse flux factors, respectively, as defined by

\begin{equation}
L_{\nu_{\rm e}}=4\pi r^{2}\frac{2\pi c}
                               {(hc)^{3}}\int dE_{\nu_{\rm e}} d\mu_{\nu_{\rm e}} E^{3}_{\nu_{\rm e}} \mu_{\nu_{\rm e}} f
\label{eq:nuelumin}
\end{equation}

\begin{equation}
\langle E^{2}_{\nu_{\rm e}}\rangle = \frac{\int dE_{\nu_{\rm e}}d\mu_{\nu_{\rm e}} E^{5}_{\nu_{\rm e}} f}
                                          {\int dE_{\nu_{\rm e}}d\mu_{\nu_{\rm e}} E^{3}_{\nu_{\rm e}} f}
\label{eq:nuerms}
\end{equation}

\begin{equation}
\langle \frac{1}{\cal F}\rangle =\frac{\int dE_{\nu_{\rm e}} d\mu_{\nu_{\rm e}} E^{3}_{\nu_{\rm e}} f}
                                     {\int dE_{\nu_{\rm e}} d\mu_{\nu_{\rm e}} E^{3}_{\nu_{\rm e}} \mu_{\nu_{\rm e}} f}
=\frac{cU_{\nu_{\rm e}}}{F_{\nu_{\rm e}}}
\label{eq:nuefluxfac}
\end{equation}
\smallskip

\noindent Corresponding quantities
are similarly defined for the electron antineutrinos.
In equations (\ref{eq:nuelumin})--(\ref{eq:nuefluxfac}),
$f$ is the electron-neutrino distribution function, which is a function
of the electron-neutrino direction cosine, $\mu_{\nu_{\rm e}}$, and energy, $E_{\nu_{\rm e}}$.
In equation (\ref{eq:nuefluxfac}), $U_{\nu_{\rm e}}$ and $F_{\nu_{\rm e}}$
are the electron-neutrino energy density and flux. 
Success in generating explosions by neutrino heating must ultimately 
rest on these three key neutrino quantities. In Figures 8 and 9, we 
plot them at a radius of 1000 km, as a function of time during our 
simulation. Figure 8a shows the electron neutrino and antineutrino
luminosities and {\small RMS} energies over the course of our entire
simulation, whereas Figure 9 provides more detail over the crucial 
first 100 ms. The inverse flux factor is not plotted here because 
it is constant and unity at this radius. The luminosities and {\small
RMS} energies at $r=$1000 km are representative of the corresponding
values in the gain region because they flatten out above the neutrinosphere 
radii. (For completeness, Figure 8b shows the muon and tau neutrino and 
antineutrino luminosities and {\small RMS} energies over the same period.)

Figure 10a gives the angle-averaged heating rate per gram as a function of radius at $t_{\rm pb}=137, 
212, 512$ ms. The gain radius moves inward during this time, starting out at a radius 
of about 70 km and finishing up at a radius of about 55 km. Note also that as the 
gain radius moves inward, the shock also recedes, and the portion of the 
curve for which $\dot{\epsilon}$ is positive extends with time over a smaller radial 
region, \ie\ , the gain volume decreases with time. 

In Figure 10b we plot the maximum net heating and net cooling rates as a
function of time during the course of our simulation. Figure 10c provides more
detail over the first 150 ms. The drop and flattening in our net heating rate
between the start of our run and 100 ms is consistent with the behavior 
exhibited by Janka and M\"{u}ller (1996) in their Figure 13 for their
exploding {\small 2D} model ``T4c.'' Between 50--100 ms, our net heating rate 
levels off at a value $\sim$ 87 MeV/baryon/s, and the Janka and 
M\"{u}ller rate levels off at a comparable though somewhat larger
value $\sim$ 93 MeV/baryon/s. However, the most dramatic difference between our results and theirs 
surfaces when comparing the magnitudes of the net cooling rate. Between
50--100 ms, our rate is $\sim$ -193 MeV/baryon/s, whereas the Janka and 
M\"{u}ller rate is 2--3 times less: $\sim$ -75 MeV/baryon/s. 
Our dramatic increase in net cooling below the gain radius, resulting 
from our use of {\small MGFLD}, must account in 
large part for the lack of an explosion in our model despite the somewhat
comparable net heating rate in the gain region, although Janka and
M\"{u}ller also begin with a larger initial shock radius that results
from starting their Newtonian simulation with one of our general 
relativistic postbounce models. 

One final note: Equation (\ref{eq:heatrate}) is appropriate for neutrino emission and 
absorption. In our simulation, the heating contributions from neutrino--electron scattering 
({\small NES}) are negligible, amounting to 3--5\% corrections for our postshock entropies 
($\leq 17-18$). At typical postshock densities between $10^{8}$ and $10^{9}$ g/cm$^{3}$, 
entropies $\sim 30$ (almost twice as large as our entropies) would be required before 
the number density of pairs would become comparable to the baryon number density, \ie\ , 
before our {\small NES} heating contributions would double. 

Figure 11 shows the evolution of the gain and shock radii in our one- and two-dimensional
simulations. The shock follows the typical trajectory seen in all of our models: It moves
out in radius to about 200 km in a quasihydrostatic way because of the rapid 
decline in the accretion ram pressure, and then recedes with some 
oscillation (most noticeably in the one-dimensional simulation) to a radius between 100--125 km at
$t_{\rm pb}=500$ ms. Initially, the gain and shock radii diverge, giving rise to an 
increasingly larger gain region, but that trend reverses at $t_{\rm pb}\approx 150$ ms. 
Note that, as a result of convection, the gain volume in our {\small 2D} simulation is larger
than it is in our {\small 1D} simulation, but despite the increased gain volume, an explosion is 
not obtained.

More important, in Figures 12a and 12b, we plot the mass and internal energy in the
gain region as a function of time during the course of our simulation. The initial
rise in both mass and internal energy results primarily from the early dramatic
increase in the gain volume, as the shock and gain radii diverge. However, at later
times, both the gain region mass and internal energy decrease monotonically with
time. This results because (1) the gain region is decreasing in size, and (2)
the density ahead of the shock is falling off. The general trends exhibited
by Figures 12a and 12b are indicative of a flow that passes through, rather
than accumulates in, the gain region. Moreover, the ratio of the internal
energy and mass in the gain region is roughly constant as a function of 
time, indicating that the total contributions to $\dot{\epsilon}$ in this
region, from neutrino heating and cooling and from advection, cancel out. 

\newpage
\section{\bf Assessing Our Use of Precalculated 1D Neutrino Distributions}

Our results depend in part on the assumption that our electron neutrino 
and antineutrino sources remain to a good approximation spherically 
symmetric during the course of our two-dimensional run. This requires 
that there be no significant convection in the region encompassing or 
below the neutrinospheres and no significant influence of neutrino-driven
convection below the gain radii:

(A) {\it Convection Below the Neutrinospheres:}~ 
In a previous paper (Mezzacappa \etal\ 1997a), we presented compelling
evidence that, in the presence of neutrino transport, the convective transport 
of heat and leptons below the neutrinospheres by protoneutron star convection 
will be significantly reduced. Our numerical results were supported by 
timescale analyses and by a simple analytical model, although final conclusions
regarding the extent of protoneutron star convection await fully
self-consistent multidimensional multigroup radiation hydrodynamics simulations.
Nonetheless, these results are mentioned here in support of the conclusions reached in 
this paper. In the absence of significant protoneutron star convection, the 
imposition of a one-dimensional spherically-symmetric neutrino radiation 
field in the region between the neutrinospheres and the shock, 
used to compute the neutrino heating and cooling there, should be a good approximation. 

(B) {\it The Influence of Neutrino-Driven Convection Below the Gain Radii:}~ 
Because our current prescription does not implement a self-consistent two-dimensional 
radiation hydrodynamics solution, we cannot capture enhancements in the neutrino 
luminosities emanating from the neutrinosphere region that result from (1) 
non--spherically-symmetric accretion through the gain radius and/or (2) inwardly 
propagating nonlinear waves that compress and heat the neutrinosphere region in a 
non--spherically-symmetric way. For example, the dense finger-like low-entropy 
inflows in the neutrino-driven convection region may penetrate the gain radius 
and strike the protoneutron star surface (Burrows \etal\ 1995, Janka \& M\"{u}ller 
1996). It has been suggested that the associated luminosity enhancements may help 
trigger explosions (Burrows \etal\ 1995), but conclusions regarding their benefit 
have been mixed (Janka \& M\"{u}ller 1996). 

To investigate whether these effects would have been important in our simulation, 
we compared our one- and two-dimensional density, temperature, and electron fraction 
snapshots at $t_{\rm pb}=212$ ms, \ie\ , at a time when neutrino-driven convection was most 
vigorous. The results are represented graphically in Figure 13. Up to the 
neutrinosphere radii ($\sim 51$ km), we found no differences. Between the neutrinospheres 
and the gain radii ($\sim 90$ km), we found hot spots in our two-dimensional simulation 
where $\Delta T/T\sim$ 3\% over $\sim$ 1/8--1/6 of the volume and $\sim$ 6\% over $\sim$
1/8--1/6 of the volume, and $Y_{\rm e}$-enhanced spots where $\Delta Y_{\rm e}/Y_{\rm e} 
\sim$ 9\% over $\sim$ 1/4--1/3 of the volume. 
However, at $t\sim 100-200$ 
ms after bounce, $L_{\nu_{\rm e}}(50\, {\rm km})\approx 2.4\times 10^{52}$ erg/s and 
$L_{\nu_{\rm e}}(90\, {\rm km})\approx 3.4\times 10^{52}$ erg/s; therefore, only 
$\sim 33\%$ of the neutrino luminosities would have been affected by these temperature 
and electron-fraction enhancements. (The percentages for electron antineutrinos are 
comparable: $\sim 50\%$ at 100 ms and $\sim 33\%$ at 200 ms.) We also considered
the enhancements to the electron neutrino and antineutrino pair emissivity due to
the hot spots, because of its strong ($T^{9}$) dependence on the local 
matter temperature. In our model the pair emissivity contributes only $\sim$10\% to
the total electron neutrino and antineutrino emissivity. The hot spots would increase 
this contribution, but only to $\sim$15\% of the total emissivity.

Considering the small local enhancements in $T$ and $Y_{\rm e}$, the small
percentage of the volume in which they occur, and the fraction of
the neutrino luminosities that would be affected by them, we do not expect 
these enhancements to have significant ramifications for the
supernova outcome.

\newpage
\section{\bf Summary, Comparisons, and Conclusions}

With two-dimensional ({\small PPM}) hydrodynamics ``coupled'' to one-dimensional
{\small MGFLD} neutrino transport, we see vigorous --- in some regions supersonic --- 
neutrino-driven convection develop behind the shock. Despite this, we do not 
obtain explosions for what should be an ``optimistic'' 15 M$_{\odot}$ model. 
Beginning with realistic postbounce initial conditions, our simulation has 
been carried out for $\sim$500 ms, a period that is long relative to the 
50--100 ms explosion timescales obtained by other groups, for models that 
explode.

An important and very interesting feature of our two-dimensional model is that, 
even in the presence of neutrino-driven convection, the angle-averaged density, 
entropy, electron fraction, and radial velocity do not differ very much from their 
counterparts in our accompanying one-dimensional {\small MGFLD} run. The differences 
arise primarily because convection in our two-dimensional simulation has moved the 
shock somewhat farther out in radius. This indicates that, while vigorous convection 
may be present, it does not contribute in our models in any significant way to the 
angle-averaged shock dynamics. 

The differences in outcome from group to group, and even from model to model for 
a given group, most likely result in large part from several factors: 
(1) Differences between numerical hydrodynamics 
methods, in particular, {\small SP} and {\small PPM} hydrodynamics, 
most likely contribute. With {\small PPM} hydrodynamics, neutrino-driven convection
is more turbulent, and consequently, the separation between high- and low-entropy
matter in the gain region, exhibited in the Herant \etal\ simulations, does not obtain. This
may have an impact on the neutrino heating efficiency, as discussed by Herant \etal\ in proposing
their ``Carnot engine'' interpretation of the supernova mechanism. (Although we again note 
that Miller \etal\ found low-mode convection, but not a ``Carnot engine.'') (2) As stressed
in Bruenn and Mezzacappa (1994), it is important to begin any {\small 2D} 
simulation with initial postbounce conditions that have been generated by
realistic core collapse and bounce, which necessitates the use of 
neutrino transport that includes all relevant neutrino interactions, particularly
neutrino-electron scattering ({\small NES}). {\small NES} is known to have a significant effect on
the deleptonization of the core during infall, and consequently, on the 
location and strength of the shock after bounce.  We begin our simulations
with postbounce configurations that have been evolved through collapse and
bounce with {\small MGFLD}, and {\small NES} included. 
In Figure 5a of Burrows \etal\ (1995), the early shock trajectory behaves like 
an initially stronger ``prompt'' shock. In contrast, in our simulations the 
shock moves out quasihydrostatically as a result of the accretion of 
matter through it. The greater initial shock strength in the Burrows \etal\ 
simulation most likely results from the difference between simulating core collapse 
with a neutrino leakage scheme (Burrows and Lattimer 1986), as opposed to {\small MGFLD} 
(Bruenn 1985). 
For the simulations presented in Herant et al. (1994), the
same gray transport scheme is used both for their {\small 1D} core collapse
evolution and their {\small 2D} postbounce evolution. It would be enlightening
to investigate the differences in postbounce conditions, and in
particular, shock location and strength, obtained using gray and
multigroup transport during core collapse and bounce. 
It is also important to stress 
that we use a {\small 1D} Newtonian postbounce profile to start our {\small 2D} Newtonian 
simulations. For Janka and M\"{u}ller (1996), the initial shock radius
increased from $\sim 120$ km to $\sim 200$ km, \ie\ , by nearly a 
factor of 2, because they started their {\small 2D} Newtonian simulation with
our {\small 1D} general relativistic postbounce model. This initial boost in shock radius
will have an effect on the subsequent shock trajectory. 
(3) Ultimately, success or failure in generating supernova explosions rests on the
right combination of neutrino luminosities, {\small RMS} energies, and inverse flux 
factors.  Whereas it is possible to list a number of inputs that might differ 
from group to group, and within each group, from simulation to simulation, the 
one that stands out the most is the neutrino {\small RMS} energy. Of course, 
the neutrino heating rate depends on the square of the neutrino {\small RMS} 
energy; therefore, differences in this quantity are magnified when folded 
into the final heating rate. For example, we find that with the Burrows \etal\ 
(1995) specification of both $T_{\nu_{\rm e}}$ and $\eta_{\nu_{\rm e}}$, 
the relationship between their neutrinosphere temperature and neutrino {\small RMS} 
energy for their exploding ``star'' model is $<E_{\nu_{\rm e}}^{2}>^{1/2}=
3.6T_{\nu_{\rm e}}$. When we fit our electron neutrino spectrum at our neutrinosphere 
radius, we obtain a characteristic dependence on temperature of $3.0T_{\nu_{\rm e}}$. 
This difference translates to a 40--50\% increase in the neutrino heating rate for 
the Burrows \etal\ ``star'' model, with commensurate ramifications for generating 
explosions. Burrows \etal\ note that explosions are not obtained in other models, 
presumably when the values of $T_{\nu_{\rm e}}$ and $\eta_{\nu_{e}}$ are chosen 
differently. Similarly, our mean electron neutrino and antineutrino energies, 
relative to those given by Herant \etal\ (1994) for their 25 M$_{\odot}$ model 
in their Figure 10, are significantly 
lower. At 100 ms after bounce, our mean electron neutrino and antineutrino 
energies are 10 and 13 MeV, respectively, compared with the 13--14 and $\sim$ 
20 MeV values obtained by Herant \etal\ The differences between our results
and the results obtained by the two groups mentioned above result from: (1) 
specification (Burrows \etal\ ) rather than computation 
of the neutrino spectra in optically thin regions; (2) patching 
together optically thick and thin regions, rather than having a transport 
scheme that transits through both regions. The maximum 
net heating rates obtained by Janka and M\"{u}ller (1996) between their gain 
radii and shock are comparable to but somewhat higher than ours: $\sim$ 93 
MeV/baryon/s, versus $\sim$ 87 MeV/baryon/s; however, their maximum net cooling 
rates below their gain radii are 2--3 times lower than ours: 
$\sim$ -75 MeV/baryon/s, versus $\sim$ -193 MeV/baryon/s. The {\small MGFLD} 
treatment we use gives a much greater neutrino cooling rate in deeper regions, 
which, as is well known, undermines efforts to generate explosions (\eg\ , see 
Herant \etal\ 1992). 

We have computed the mass and internal energy in the gain region as a
function of time to address the issue of whether or not neutrino-driven
convection leads to greater neutrino heating efficiency and the accumulation
of mass and energy in the gain region [Bethe (1990), Herant \etal\ (1994)]. 
Other interpretations suggest that supernovae result as critical phenomena  
when the neutrino luminosities are sufficiently high, given the ram pressure
of the preshock matter, to render the flow unstable to explosion [Burrows
and Goshy (1993), Burrows \etal\ (1995)]. We find an increase in mass and
internal energy in the gain radius up to about 150 ms after bounce, as a
result of the increasing gain volume as the gain and shock radii diverge.
After that, we find a monotonic decrease in both quantities, consistent
with the density falloff in the preshock matter and with matter flowing 
through, rather than accumulating in, the gain region.

To assess our use of precalculated {\small 1D} neutrino distributions for matter
heating and deleptonization, we have considered the non--spherically-symmetric 
luminosity enhancements
that would occur from local temperature and electron fraction enhancements 
below the gain radii (which enclose the electron neutrino and antineutrino 
sources) that are seen in our two-dimensional run, which result either from 
non--spherically-symmetric accretion through the gain radius or nonlinear 
inwardly propagating non--spherically-symmetric waves. We see no enhancements 
below the neutrinosphere radii; between them and the gain radii, we see small 
enhancements that occur over a small fraction of the volume responsible 
for producing less than 1/3 of the neutrino luminosities. Therefore,
we do not expect these enhancements to have dynamical consequences. 

We do not expect to obtain explosions for more massive stars. Moreover, our 
simulations are Newtonian. With general relativistic gravity, conditions 
will be even more pessimistic. The neutrino luminosities will be redshifted,
the increased infall velocities and the smaller width between the gain radii 
and the shock will allow less time for neutrino heating to reverse infall, 
and everything will occur in a deeper gravitational well, making explosion
more difficult.
  
We are in the process of carrying out
simulations with ray-by-ray {\small MGFLD} coupled to two-dimensional hydrodynamics
in an effort to (1) ``bracket'' the approximation of using one-dimensional
neutrino transport in a two-dimensional setting and (2) step toward a two-dimensional
multigroup neutrino transport scheme. The imposition of spherical symmetry
in our current model maximizes the lateral transport of neutrinos in regions
that are optically thick, which would have a tendency to minimize convection
in that region (Mezzacappa \etal\ 1997a); whereas ray-by-ray transport, by
definition, minimizes it. Our results regarding neutrino-driven convection
in the gain region hinge on our assumption that the neutrino radiation field
there is realistically determined by one-dimensional {\small MGFLD} neutrino
transport in the region near and below the neutrinospheres. (Otherwise, our
use of {\small 1D} precalculated distributions would not be a good approximation.)
``Bracketing''
our one-dimensional neutrino transport approximations will give us a better sense of how realistic
these approximations are. Of course, 
final conclusions regarding our {\small 1D} transport approximations
await fully self-consistent two-dimensional multigroup radiation hydrodynamics
simulations.

On a more optimistic note, recently we have obtained
new results from comparisons of three-flavor Boltzmann neutrino transport
and three-flavor {\small MGFLD} in postbounce supernova environments (thermally
frozen, hydrostatic). In particular, the Boltzmann
net heating rate in the region directly above
the gain radii is significantly larger (Mezzacappa \etal\ 1997b). These
results suggest that Boltzmann transport will yield greater neutrino heating
and more vigorous neutrino-driven convection; both would increase the chances
of reviving the stalled shock. 

Finally, it is well known that in two and three dimensions energy cascades
in different directions, from short- to long-wavelength modes in two dimensions,
and in the opposite direction in three dimensions (Porter, Pouquet, \& Woodward 
1992). Consequently, we expect
three-dimensional simulations to ``look'' more like one-dimensional simulations
than do our two-dimensional simulations. Therefore, it may be more, not less, difficult to
obtain explosions in 3D, if success in generating explosions relies heavily on
convection. We are currently investigating the dependence of neutrino-driven
convection on the number of spatial dimensions (Knerr \etal\ 1997). 
To make matters worse, ultimately we will have the task of obtaining 
general relativistic 3D explosions, with or without the aid of convection, and
if one considers other complexities, like (1) the uncertainties in the precollapse
models, high-density equation of state [\eg, see Keil \& Janka (1995)], and high-density neutrino opacities
[\eg, see Raffelt and Seckel (1994)], and (2) the noninclusion of rotation [\eg, see Shimizu \etal\ (1994)] 
and other potentially important input physics in simulations
that include realistic multidimensional hydrodynamics and neutrino transport,
we are far from being able to say definitively how supernovae explode and
whether or not a single component of the problem, like convection, is the
key to unlocking it.

\newpage
\section{Acknowledgements}

AM, ACC, MWG, and MRS were supported at the Oak Ridge National 
Laboratory, which is managed by Lockheed Martin Energy Research 
Corporation under DOE contract DE-AC05-96OR22464. AM, MWG, and 
MRS were supported at the University of Tennessee under DOE contract 
DE-FG05-93ER40770. ACC and SU were supported at Vanderbilt 
University under DOE contract DE-FG02-96ER40975. SWB was 
supported at Florida Atlantic University under NSF grant 
AST--9618423 and NASA grant NRA-96-04-GSFC-073, and JMB 
was supported at North Carolina State University under 
NASA grant NAG5-2844. 
The simulations presented in this Letter were carried out on
the Cray C90 at the National Energy Research Supercomputer
Center, the Cray Y/MP at the North Carolina Supercomputer 
Center, and the Cray Y/MP and Silicon Graphics Power
Challenge at the Florida Supercomputer Center. We would 
like to thank 
Willy Benz, 
Adam Burrows,
Chris Fryer, 
Wolfgang Hillebrandt,
Thomas Janka, 
Ewald M\"{u}ller,
Michael Smith, 
Doug Swesty,
and 
Friedel Thielemann 
for stimulating discussions; and especially the referee, Stirling
Colgate, for many important comments, questions, and suggestions
that improved the content of this paper significantly. 

\newpage
\section{\bf References}

\noindent Arnett, W. D. 1986, in IAU Symposium 125, The Origin and Evolution of
     Neutron Stars,\linebreak\indent ed. D. J. Helfand and J. H. Huang (Dordrecht: Reidel)

\noindent Arnett, W. D. 1987, ApJ, 319, 136

\noindent Arnett, W. D., Fryxell, B., \& M\"{u}ller, E. 1989, ApJ, 341, L63

\noindent Bethe, H. A. 1990, Rev. Mod. Phys., 62, 801

\noindent Bethe, H. A. 1993, ApJ, 412, 192

\noindent Bethe, H. A., Brown, G. E., \& Cooperstein, J. 1987, ApJ, 322, 201

\noindent Bethe, H. \& Wilson, J. R. 1985, ApJ, 295, 14

\noindent Bowers, R. L., \& Wilson, J. R. 1991, Numerical Modeling in Applied
      Physics and\linebreak\indent Astrophysics (Boston: Jones \& Bartlett), chap. 5

\noindent Bruenn, S. W. 1993, in Nuclear Physics in the Universe, eds. M. W.
      Guidry and\linebreak\indent M. R. Strayer (IOP Publishing, Bristol), p. 31

\noindent Bruenn, S. W. \& Mezzacappa, A. 1994, ApJ, 433, L45

\noindent Bruenn, S. W., Mezzacappa, A., \& Dineva, T. 1995, Phys. Rep., 256, 69

\noindent Bruenn, S. W. \& Dineva, T. 1996, ApJ, 458, L71

\noindent Burrows, A. 1987, ApJ, 318, L63

\noindent Burrows, A. \& Lattimer, J. M. 1988, Phys. Rep., 163, 5

\noindent Burrows, A., Hayes, J., \& Fryxell, B. A. 1995, ApJ, 450, 830

\noindent Burrows, A. \& Fryxell, B. A. 1992, Science, 258, 430

\noindent Burrows, A. \& Fryxell, B. A. 1993, ApJ, 418, L33

\noindent Colella, P., \& Woodward, P. 1994, J. Comp. Phys., 54, 174

\noindent Colgate, S. A., \& White, R. H. 1966, Ap. J., 143, 626

\noindent Colgate, S. A., Herant, M. E., \& Benz, W. 1993, Phys. Rep., 227, 157

\noindent Cook, W. R., \etal\ 1988, ApJ, 334, L87

\noindent Cooperstein, J. 1993, in Nuclear Physics in the Universe, eds. M. W.
      Guidry and\linebreak\indent M. R. Strayer (IOP Publishing, Bristol), p. 99

\noindent Den, M., Yoshida, T., \& Yamada, Y. 1990, Progr. Theor. Phys., 83, 723

\noindent Dotani, T., \etal\ 1987, Nature, 330, 230

\noindent Epstein, R. I. 1979, MNRAS, 188, 305

\noindent Erikson, E. F., Hass, M. R., Colgan, S. W. J., Lord, S. D., Burton, M. G.,
      Wolf, J.,\linebreak\indent Hollenbach, D. J., \& Werner, H. 1988, ApJ, 330, L39

\noindent Fryxell, B., Arnett, W. D., \& M\"{u}ller, E. 1991, ApJ, 367, 619

\noindent Gehrels, N., MacCallum, C. J., \& Leventhal, M. 1987, ApJ, 320, L19

\noindent Gehrels, N., Leventhal, M., \& MacCallum, C. J. 1988, in Nuclear
      Spectroscopy of\linebreak\indent Astrophysical Sources, ed. N. Gehrels \& G. Share
      (New York: AIP), p. 87

\noindent Haas, M. R., Colgan, S. W. J., Erickson, E. F., Lord, S. D., Burton, M. G.,
     \&\linebreak\indent Hollenbach, D. J. 1990, ApJ, 360, 257

\noindent Hachisu, I., Matsuda, T., Nomoto, K., Shigeyama, T. 1990, ApJ, 358, L57 

\noindent Herant, M., \& Benz, W. 1991, ApJ, 370, L81

\noindent Herant, M., \& Benz, W. 1992, ApJ, 387, 294

\noindent Herant, M., Benz, W., \& Colgate, S. A. 1992, ApJ, 395, 642

\noindent Herant, M., Benz, W., Hix, W. R., Fryer, C. L., \& Colgate, S. A. 1994,
      ApJ, 435, 339

\noindent H\"{o}flich, O. 1988, in IAU Colloq. 108, Atmospheric Diagnostics of Stellar
      Evolution,\linebreak\indent ed. K. Nomoto (Berlin: Springer), p. 288

\noindent Janka, H.-Th. \& M\"{u}ller, E. 1993a, in Proc. of the IAU Coll.
      145 (Xian China,\linebreak\indent May 24-29, 1993), eds. R. McCray and Wang Zhenru,
      (Cambridge Univ. Press,\linebreak\indent Cambridge); MPA-Preprint 748

\noindent Janka, H.-Th. \& M\"{u}ller, E. 1993b, in Frontiers of Neutrino
      Astrophysics, eds. Y. Suzuki\linebreak\indent and K. Nakamura, (Universal Academy Press,
      Tokyo). p. 203

\noindent Janka, H.-Th. \& M\"{u}ller, E. 1995, Ap. J., 448, L109

\noindent Janka, H.-Th., \& M\"{u}ller, E. 1996, A\&A 306, 167

\noindent Keil, W. \& Janka, H.-Th. 1995, A\&A, 296, 145

\noindent Keil, W., Janka, H.-Th, \& M\"{u}ller, E. 1996, ApJ, 473, L111

\noindent Knerr, J., Blondin, J. M., Mezzacappa, A., and Bruenn, S. W. 1997, in preparation

\noindent Mahoney, W. A., \etal\ 1988, ApJ, 334, L81

\noindent Matz, S. M., Share, \& Chupp, E. L., 1988, in Nuclear
      Spectroscopy of Astrophysical\linebreak\indent Sources, ed. N. Gehrels \& G. Share
      (New York: AIP)

\noindent Matz, S. M., Share, G. H., Leising, M. D., Chupp, E. L., Vestrand, W. T.,
      Purcell, W. R.,\linebreak\indent Strickman, M. S., \& Reppin, C. 1988, Nature, 331, 416

\noindent Mayle, R. W. 1985, Ph.D. Thesis, Univ. California, Berkeley (UCRL preprint
      no. 53713)

\noindent McCray, R., Shull, J. M., \& Sutherland, P. 1988, ApJ, 327, L73

\noindent Mezzacappa, A., Messer, O. E. B., Bruenn, S. W., \& Guidry, M. W. 1997b, in preparation 

\noindent Mezzacappa, A., Calder, A. C., Bruenn, S. W., Blondin, J. M.,
      Guidry, M. W.,\linebreak\indent Strayer, M. R., \& Umar, A. S. 1997a, ApJ, in press

\noindent Miller, D. S., Wilson, J. R., \& Mayle, R. W. 1993, ApJ, 415, 278

\noindent M\"{u}ller, E. 1993, in Proc. of the 7th Workshop on Nuclear Astrophysics
      (Ringberg\linebreak\indent Castle, March 22-27, 1993), eds. W. Hillebrandt and E.
      M\"{u}ller, Report MPA/P7,\linebreak\indent Max-Plank-Institut f\"{u}r Astrophysik,
      Garching, p. 27

\noindent M\"{u}ller, E. \& Janka, H.-Th. 1994, in Reviews in Modern Astronomy 7
     (Proc. of the Int.\linebreak\indent Conf. of the AG (Bochum, Germany, 1993)), eds. G. Klure
     (Astronomische Gesellschaft,\linebreak\indent Hamburg), p. 103

\noindent M\"{u}ller, E., Fryxell, B., \& Arnett W. D. 1991, in Proc. Elba
      Workshop on Chemical and Dynamical Evolution of Galaxies, ed. F. Federini,
      J. Franco, \& F. Matteucci (Pisa: ETS\linebreak\indent Editrice), p. 394

\noindent Pinto, P. A., \& Woosley, S. E. 1988, Nature, 333, 534

\noindent Porter, D. H., Pouquet, A., \& Woodward, P. R. 1992, Theor. Comput. Fluid Dynamics,\linebreak\indent 4, 13

\noindent Raffelt, G., \& Seckel D. 1994, Phys. Rev. D, to be published

\noindent Rank, D. M., \etal\ 1988, Nature, 331, 505

\noindent Sandie, W. G., \etal\ 1988, ApJ, 334, L91

\noindent Shimizu, T., Yamada, S., \& Sato, K. 1994, ApJ, 432, L119

\noindent Spyromilio, J., Meikle, W. P. S., \& Allen, D. A. 1990, MNRAS, 242, 669

\noindent Sunyaev, R., \etal\ Nature, 330, 227

\noindent Teegarden, B. J., \etal\ 1989, Nature, 339, 122

\noindent Tuller, J., Barthelemmy, S., Gehrels, N., Teegarden, B. J., Leventhal, M.,
      \&\linebreak\indent MacCallum, C. J. 1990, ApJ, 351, L41

\noindent Weaver, T. A. \& Woosley, S. E. 1997, ApJ, in preparation 

\noindent Wilson, J. R. 1985, in Numerical Astrophysics, eds. J. M. Centrella, J. M.
      LeBlanc, \&\linebreak\indent R. L. Bowers (Boston: Jones \& Bartlett), p. 422

\noindent Wilson, J. R. \& Mayle, R. W. 1988, Phys. Rep., 163, 63

\noindent Wilson, J. R. \& Mayle, R. W. 1993, Phys. Rep., 227, 97

\noindent Wilson, R. B., \etal\ 1988, in Nuclear Spectroscopy of Astrophysical Sources,
      ed. N. Gehrels \&\linebreak\indent G. Share (New York: AIP), p. 66

\noindent Woosley, S. E. \& Weaver, T. A. 1995, ApJS, 101, 181 

\noindent Xu, Y., Sutherland, P., McCray, R., \& Ross, R. 1988, ApJ, 327, 197

\noindent Yamada, Y., Nakamura, T., \& Oohara, K. 1990, Progr. Theor. Phys., 84, 436
\end{document}


\newcounter{numbercount}
\newcounter{lettercount}
\setcounter{numbercount}{1}
\setcounter{lettercount}{1}
\setcounter{page}{42}

\renewcommand{\thefigure}{\arabic{numbercount}}
\setcounter{numbercount}{1}

\begin{figure}
\centering
\epsfig{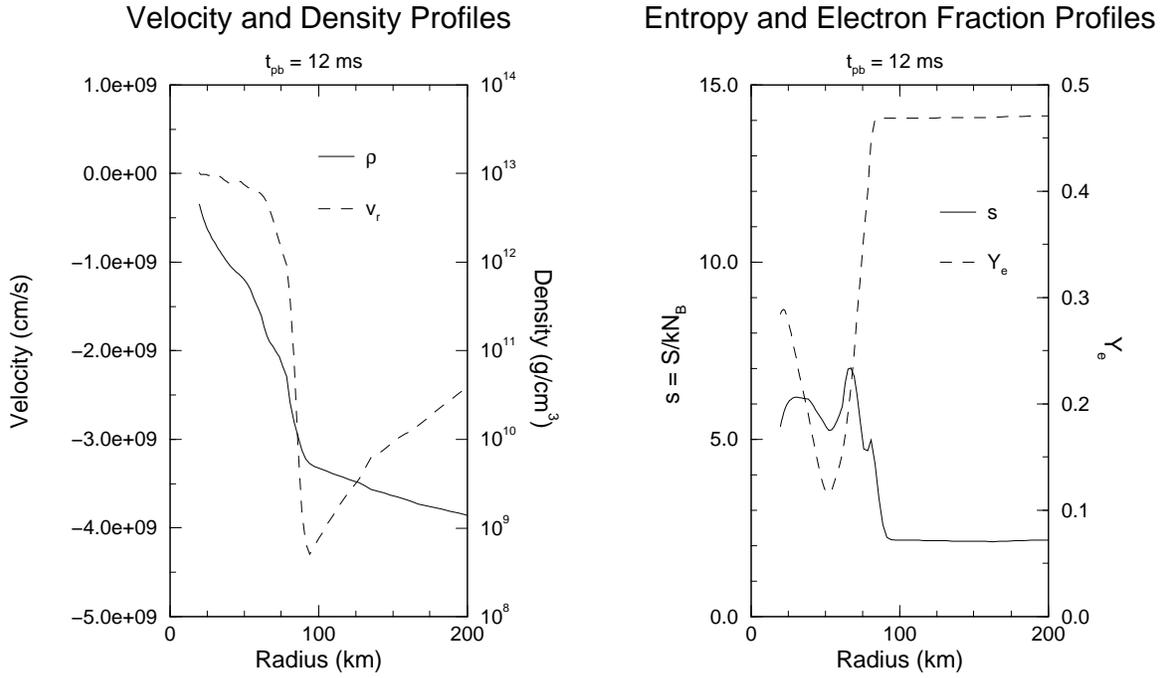}
\caption[Initial entropy and electron fraction profiles]
{\protect\small \protect\baselineskip 10pt \label{nfig1}
Initial postbounce velocity, density, entropy, and electron fraction
profiles for our 15 ${\rm M}_{\odot}$ model.
}
\end{figure}

\renewcommand{\thefigure}{\arabic{numbercount}\alph{lettercount}}
\addtocounter{numbercount}{1}
\setcounter{lettercount}{1}

\begin{figure}
\centering
\epsfig{file=nfigs/nfig2a.epsi,width=4.8in,angle=0}
\caption[Density and  entropy vs. radius]
{\protect\small \protect\baselineskip 10pt \label{nfig2a}
Plots of density and entropy versus radius
comparing one-dimensional simulations with our
{\small PPM} and {\small MGFLD} codes, at 307 ms
after the start of our run.
}
\end{figure}
\clearpage

\addtocounter{lettercount}{1}

\begin{figure}
\centering
\epsfig{file=nfigs/nfig2b.epsi,width=4.8in,angle=0}
\caption[Electron fraction and radial velocity vs. radius]
{\protect\small \protect\baselineskip 10pt \label{nfig2b}
Plots of electron fraction and radial velocity versus
radius for the same comparison shown in Figure 2a. 
}
\end{figure}

\addtocounter{lettercount}{1}

\begin{figure}
\centering
\epsfig{file=nfigs/nfig2c.epsi,width=4.8in,angle=0}
\caption[Density and  entropy vs. radius]
{\protect\small \protect\baselineskip 10pt \label{nfig2c}
Plots of density and entropy versus radius
comparing one-dimensional simulations with our
{\small PPM} and {\small MGFLD} codes, at 512 ms
after the start of our run.
}
\end{figure}
\clearpage

\addtocounter{lettercount}{1}

\begin{figure}
\centering
\epsfig{file=nfigs/nfig2d.epsi,width=4.8in,angle=0}
\caption[Electron fraction and radial velocity vs. radius]
{\protect\small \protect\baselineskip 10pt \label{nfig2d}
Plots of electron fraction and radial velocity versus
radius for the same comparison shown in Figure 2c.
}
\end{figure}

\setcounter{numbercount}{3}
\setcounter{lettercount}{2}

\begin{figure}
\centering
\epsfig{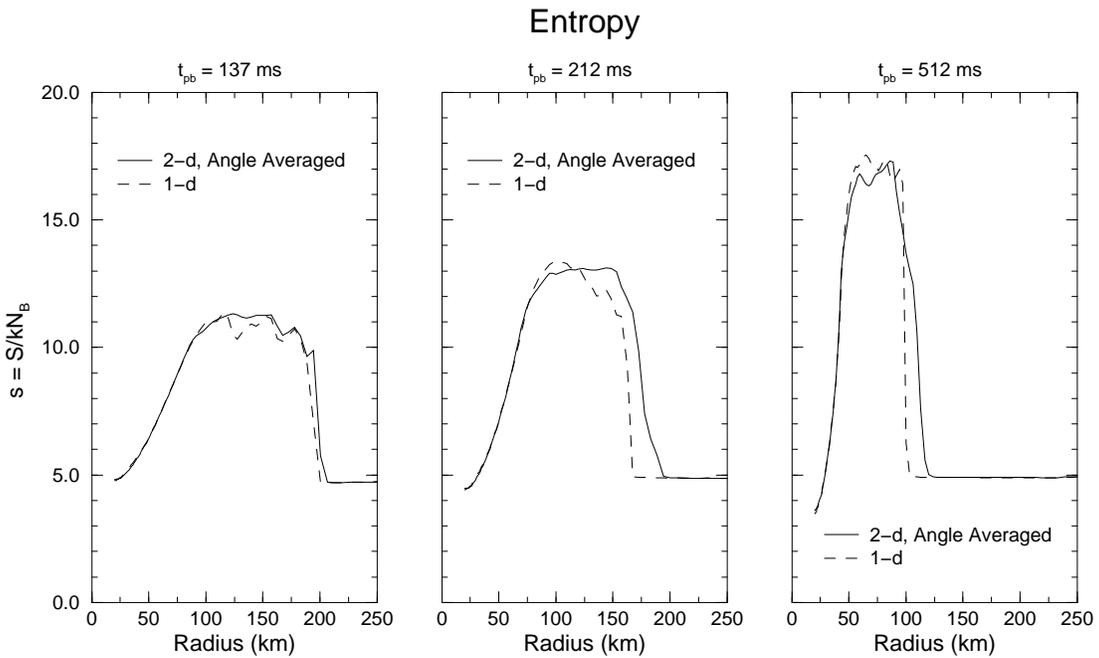}
\caption[Angle-averaged entropy]
{\protect\small \protect\baselineskip 10pt \label{nfig3b}
Plots of the angle-averaged entropy
for the three entropy snapshots shown in Figure 3a.
Also shown for comparison are the corresponding entropy
profiles from our {\small 1D PPM} simulation.
}
\end{figure}
\clearpage

\addtocounter{numbercount}{1}
\setcounter{lettercount}{2}

\begin{figure}
\centering
\epsfig{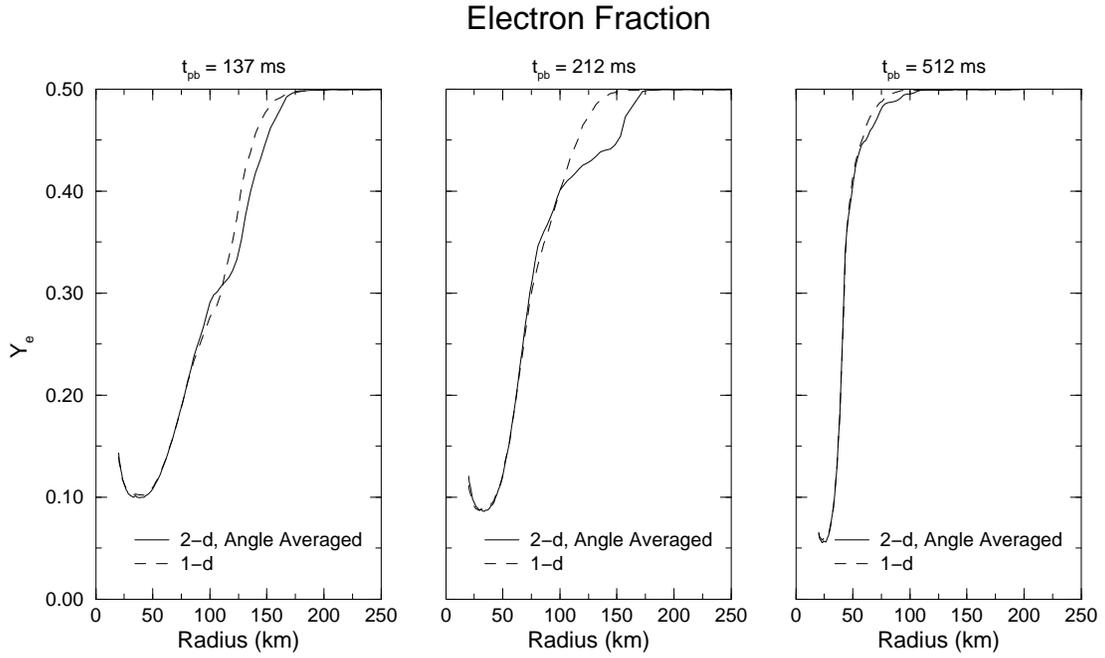}
\caption[Angle-averaged electron fraction]
{\protect\small \protect\baselineskip 10pt \label{nfig4b}
Plots of the angle-averaged electron fraction
for the three electron fraction snapshots shown in
Figure 4a. Also shown for comparison is the corresponding
electron fraction from our {\small 1D PPM} simulation. 
}
\end{figure}

\addtocounter{numbercount}{1}

\begin{figure}[p]
\centering
\epsfig{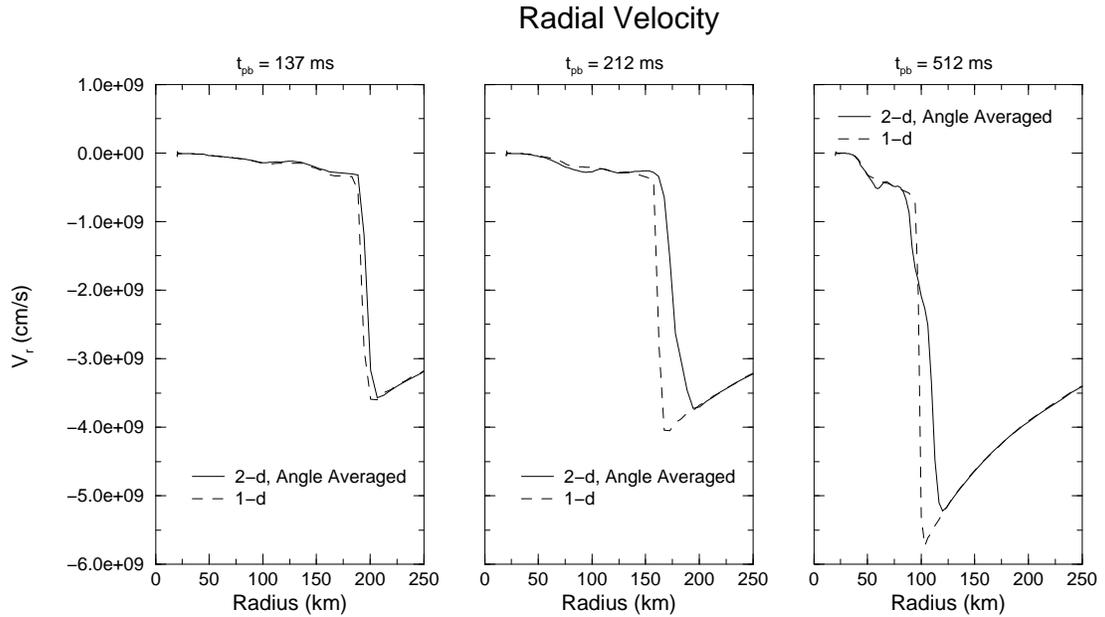}
\caption[Initial entropy and electron fraction profiles]
{\protect\small \protect\baselineskip 10pt \label{nfig5b}
Plots of the angle-averaged radial velocity
for the three radial velocity snapshots shown in
Figure 4a. Also shown for comparison is the corresponding
radial velocity from our {\small 1D PPM} simulation.
}
\end{figure}
\clearpage

\renewcommand{\thefigure}{\arabic{numbercount}}
\addtocounter{numbercount}{1}

\begin{figure}[p]
\centering
\epsfig{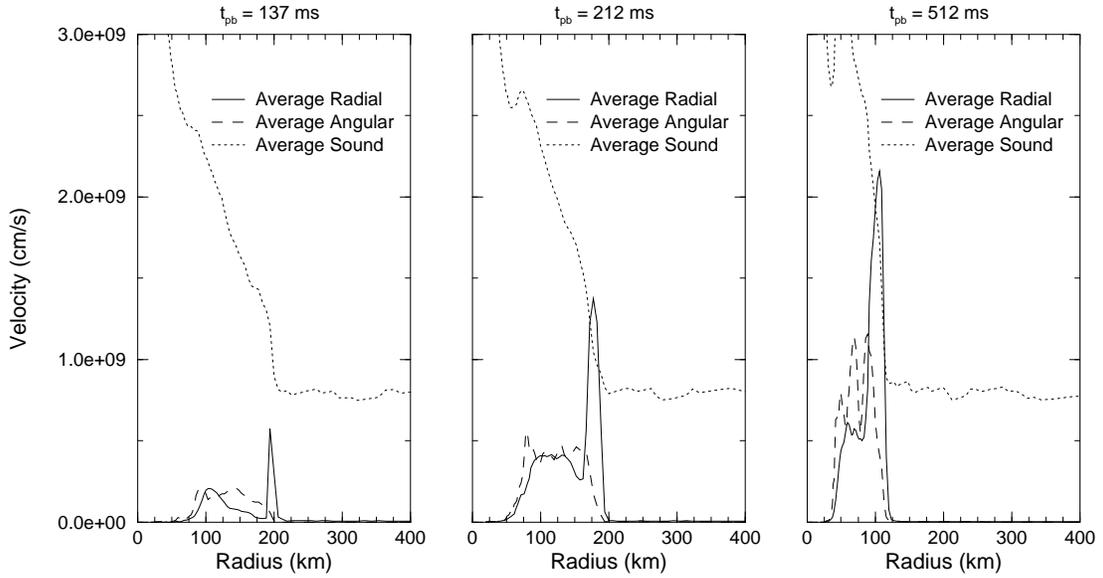}
\caption[Angle--averaged radial and angular convection velocity]
{\protect\small \protect\baselineskip 10pt \label{nfig6}
Plots of the angle-averaged radial convection velocity,
angular convection velocity, and local sound speed
at 137, 212, and 512 ms after bounce.
}
\end{figure}

\addtocounter{numbercount}{1}

\begin{figure}[p]
\centering
\epsfig{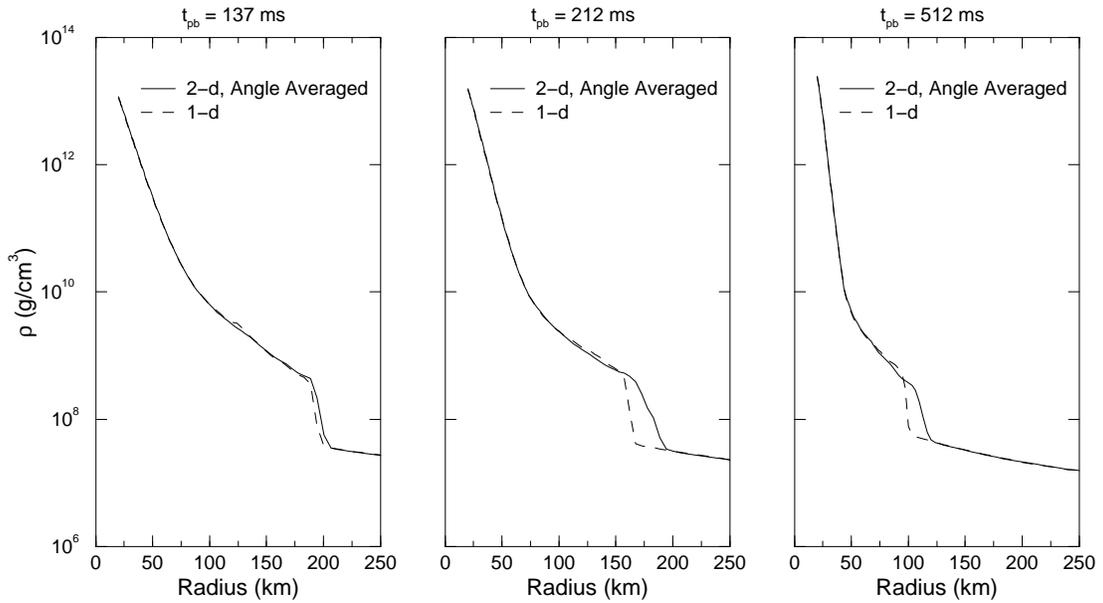}
\caption[Angle-averaged density]
{\protect\small \protect\baselineskip 10pt \label{nfig7}
Plots of the angle-averaged density
at 137, 212, and 512 ms after bounce.
Also shown for comparison is the corresponding
density from our {\small 1D PPM} simulation. 
}
\end{figure}
\clearpage

\renewcommand{\thefigure}{\arabic{numbercount}\alph{lettercount}}
\addtocounter{numbercount}{1}
\setcounter{lettercount}{1}

\begin{figure}[p]
\centering
\epsfig{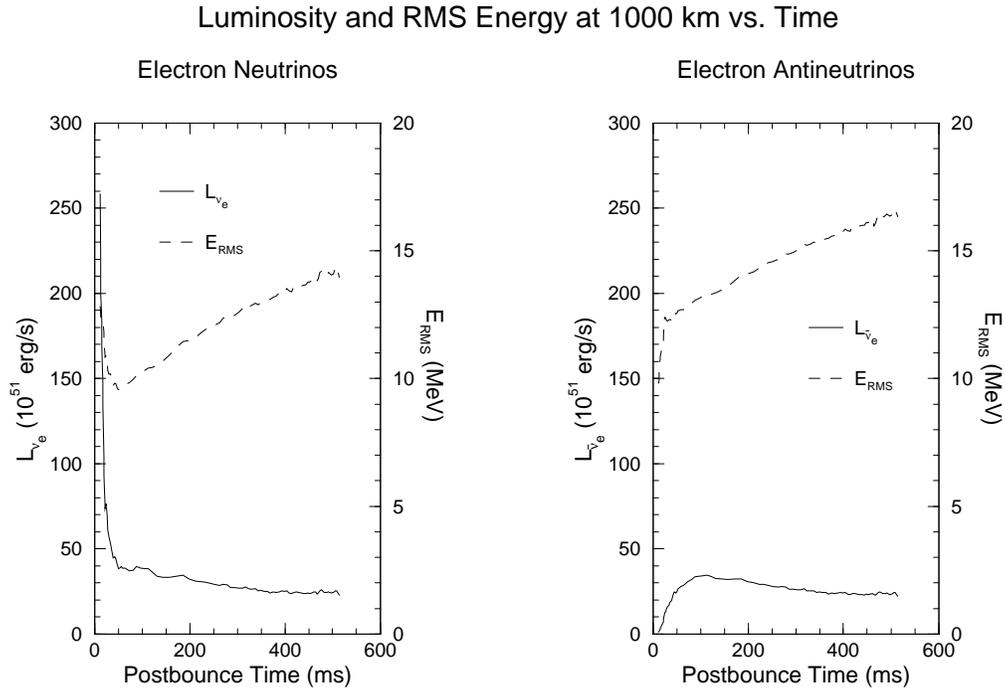}
\caption[Luminosity and RMS energy]
{\protect\small \protect\baselineskip 10pt \label{nfig8a}
Plot of the {\small MGFLD} electron neutrino and
antineutrino luminosities and RMS energies at 1000 km,
as a function of time.
}
\end{figure}

\addtocounter{lettercount}{1}

\begin{figure}[p]
\centering
\epsfig{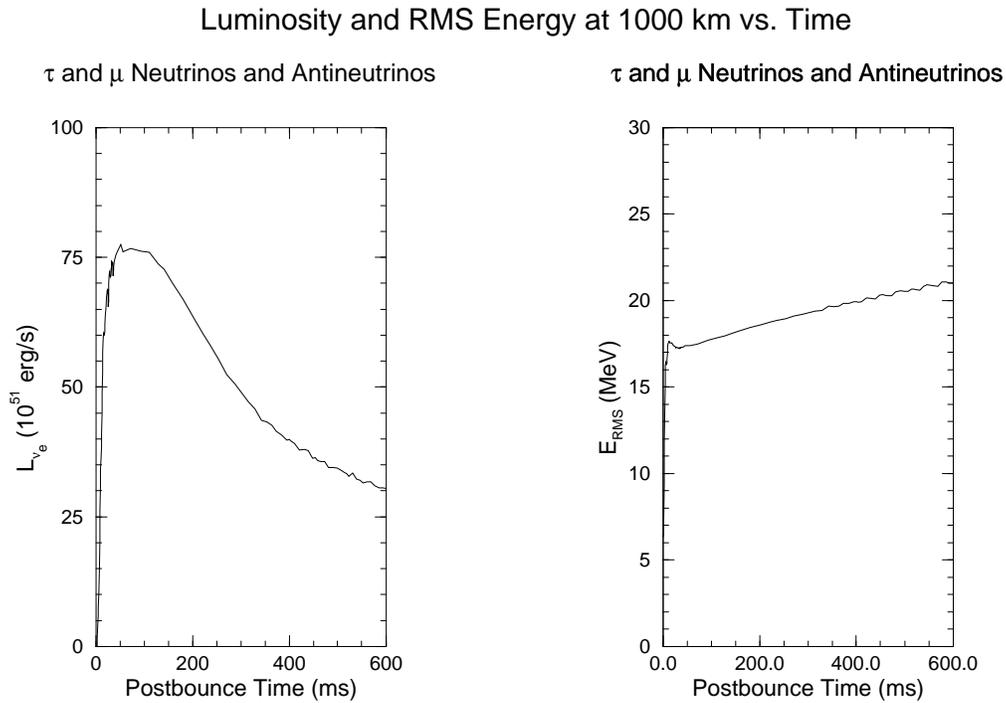}
\caption[Luminosity and RMS energy]
{\protect\small \protect\baselineskip 10pt \label{nfig8b}
Plot of the {\small MGFLD} muon and tau neutrino and
antineutrino luminosities and RMS energies at 1000 km,
as a function of time.
}
\end{figure}
\clearpage

\renewcommand{\thefigure}{\arabic{numbercount}}
\addtocounter{numbercount}{1}

\begin{figure}[p]
\centering
\epsfig{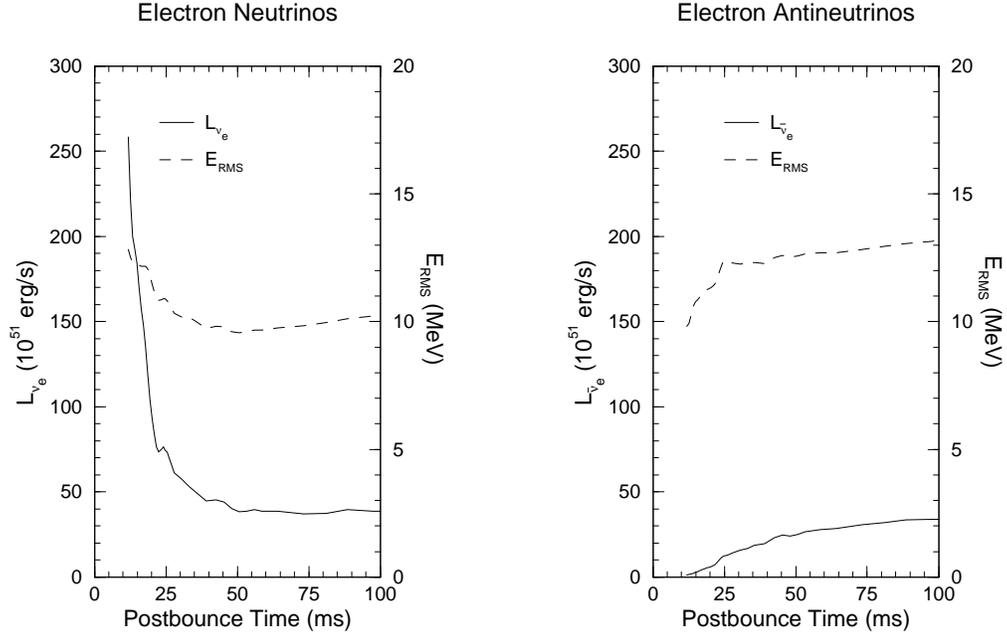}
\caption[Luminosity and RMS energy]
{\protect\small \protect\baselineskip 10pt \label{nfig9}
Early detail of the {\small MGFLD} electron neutrino and
antineutrino luminosities and RMS energies at 1000 km. 
}
\end{figure}

\renewcommand{\thefigure}{\arabic{numbercount}\alph{lettercount}}
\addtocounter{numbercount}{1}
\setcounter{lettercount}{1}

\begin{figure}
\centering
\epsfig{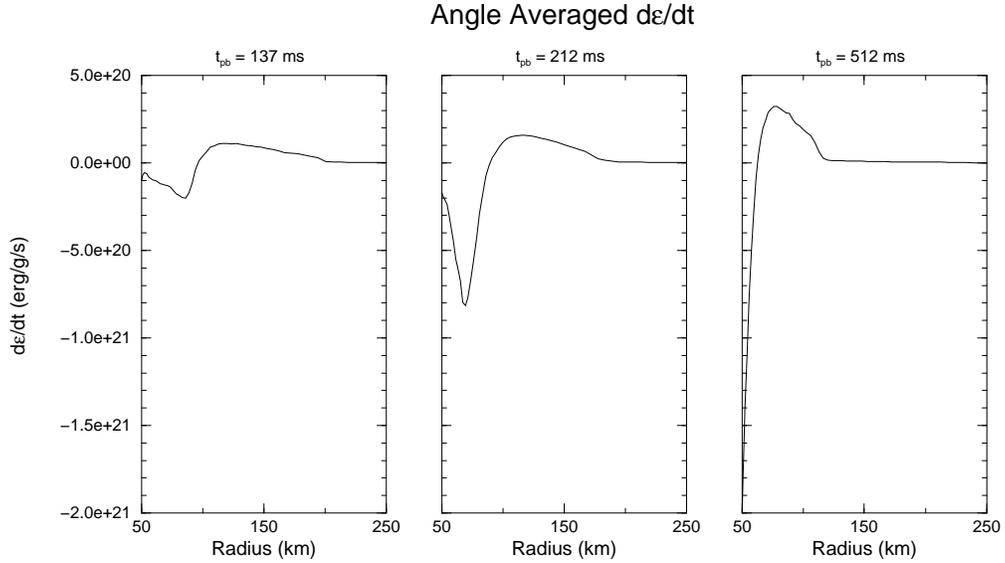}
\caption[Angle-averaged $d\epsilon/dt$]
{\protect\small \protect\baselineskip 10pt \label{nfig10a}
Plots of the angle-averaged $d\epsilon/dt$ at 137, 212, and 512
ms after bounce. $d\epsilon/dt$ is the net heating or cooling due to
electron neutrino and antineutrino absorption and emission.
}
\end{figure}
\clearpage

\addtocounter{lettercount}{1}

\begin{figure}
\centering
\epsfig{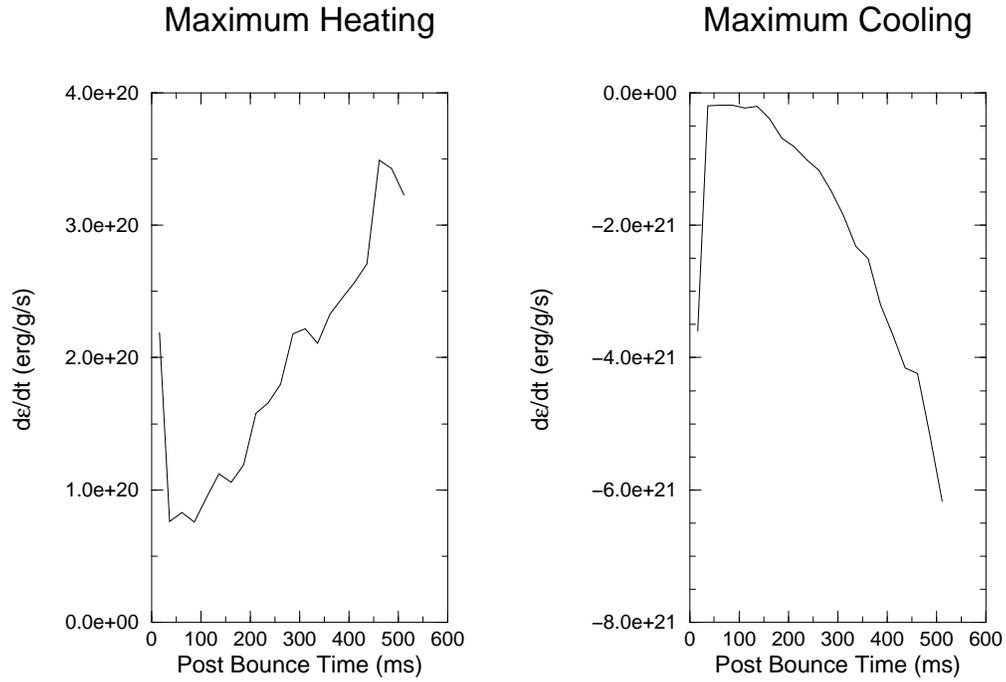}
\caption[Maximum net neutrino heating and net neutrino cooling]
{\protect\small \protect\baselineskip 10pt \label{nfig10b}
Plots of the maximum net neutrino heating and net neutrino cooling
rates as a function of time during the course of our simulation.
}
\end{figure}

\addtocounter{lettercount}{1}

\begin{figure}
\centering
\epsfig{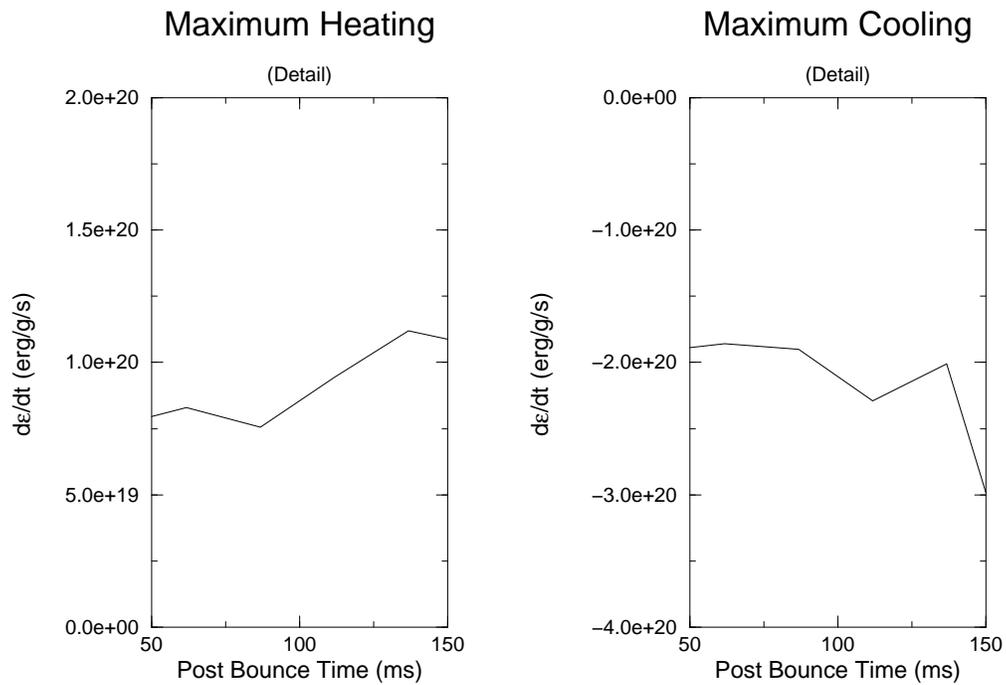}
\caption[Maximum net neutrino heating and net neutrino cooling (detail)]
{\protect\small \protect\baselineskip 10pt \label{nfig10c}
Plots of the maximum net neutrino heating and net neutrino cooling
rates as a function of time during the critical 50--150 ms after bounce.
}
\end{figure}
\clearpage

\renewcommand{\thefigure}{\arabic{numbercount}}
\addtocounter{numbercount}{1}

\begin{figure}
\centering
\epsfig{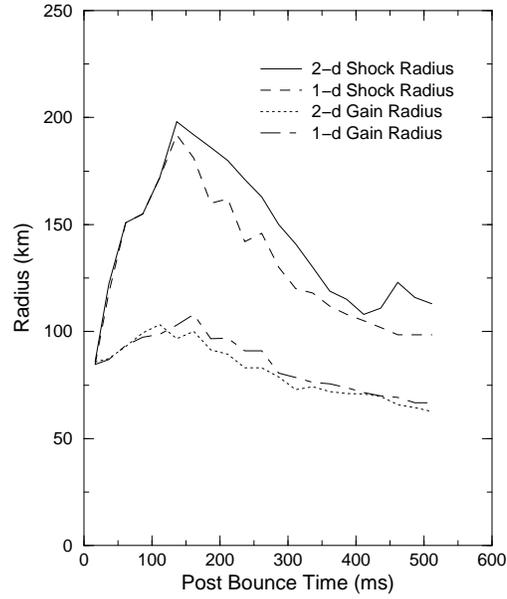}
\caption[Gain and shock radii vs. time]
{\protect\small \protect\baselineskip 10pt \label{nfig11}
Plots of the angle-averaged gain and shock radii vs. time for our
two-dimensional {\small PPM} simulation.
Also shown
for comparison are the gain and shock radii from our
{\small 1D PPM} simulation. 
}
\end{figure}

\renewcommand{\thefigure}{\arabic{numbercount}\alph{lettercount}}
\addtocounter{numbercount}{1}
\setcounter{lettercount}{1}

\begin{figure}
\centering
\epsfig{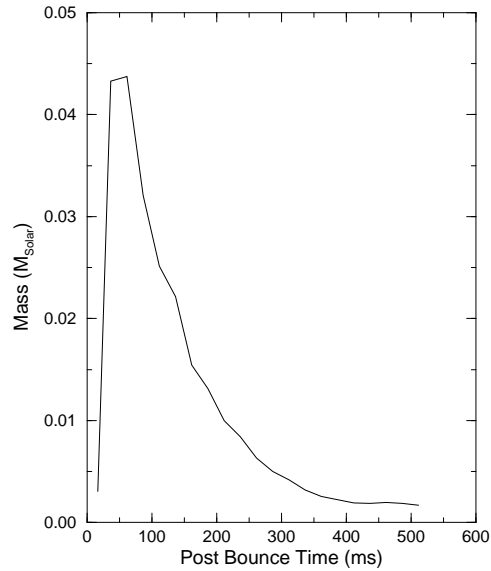}
\caption[Enclosed mass]
{\protect\small \protect\baselineskip 10pt \label{nfig12a}
Plot of gain region mass vs. time for our
two-dimensional {\small PPM} simulation. 
}
\end{figure}
\clearpage

\begin{figure}
\centering
\addtocounter{lettercount}{1}
\epsfig{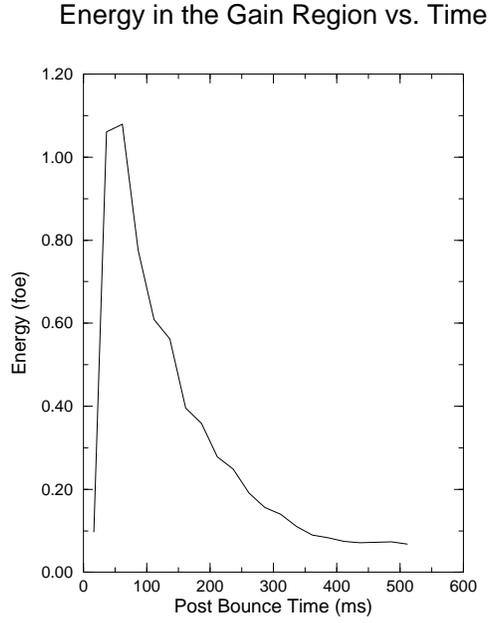}
\caption[Enclosed energy]
{\protect\small \protect\baselineskip 10pt \label{nfig12b}
Plot of gain region internal energy vs. time for
our two-dimensional {\small PPM} simulation.
}
\end{figure}

\setcounter{numbercount}{3}
\setcounter{lettercount}{1}

\begin{figure}
\centering
\epsfig{file=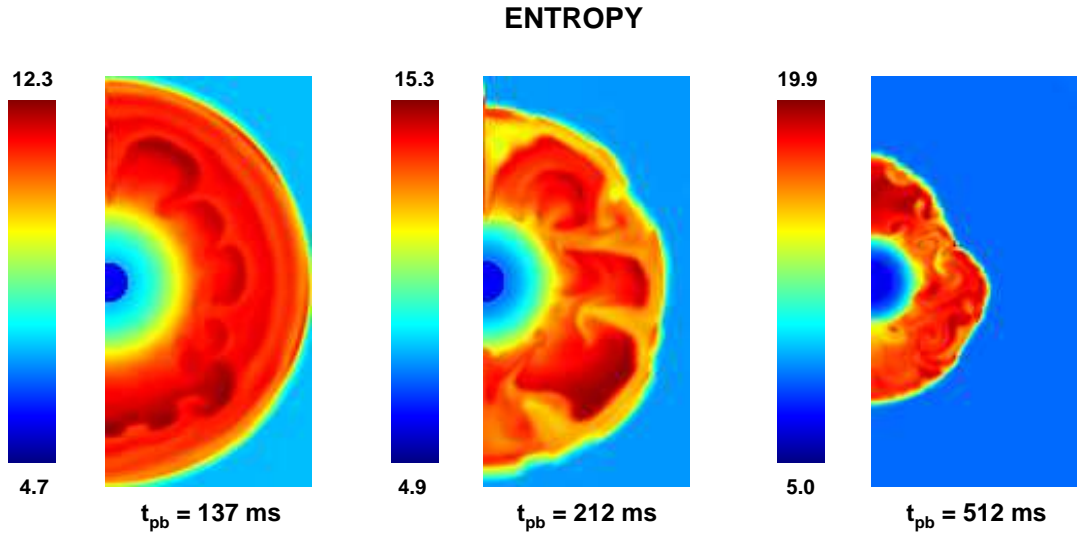,width=5.7in,angle=0}
\caption[Two-dimensional entropy]
{\protect\small \protect\baselineskip 10pt \label{nfig3a}
Two-dimensional entropy plots showing the evolution of
neutrino-driven convection in our 15 ${\rm M}_{\odot}$ model,
at 137, 212, and 512 ms after bounce.
}
\end{figure}
\clearpage

\addtocounter{numbercount}{1}

\begin{figure}
\centering
\epsfig{file=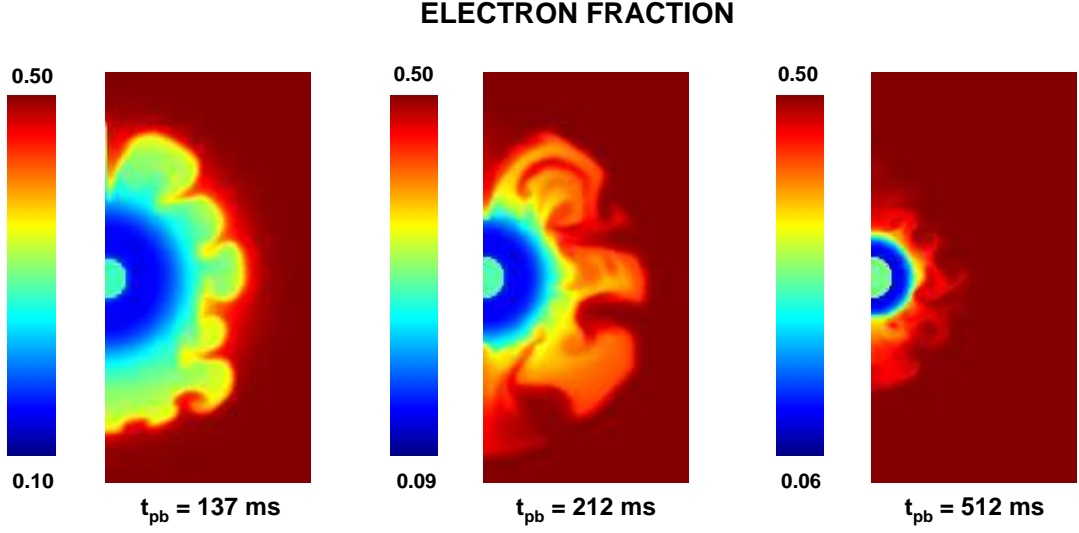,width=5.7in,angle=0}
\caption[Two-dimensional electron fraction]
{\protect\small \protect\baselineskip 10pt \label{nfig4a}
Two-dimensional electron fraction plots showing the
evolution of neutrino-driven convection in our 15 ${\rm M}_{\odot}$
model, at 137, 212, and 512 ms after bounce.
}
\end{figure}

\addtocounter{numbercount}{1}

\begin{figure}
\centering
\epsfig{file=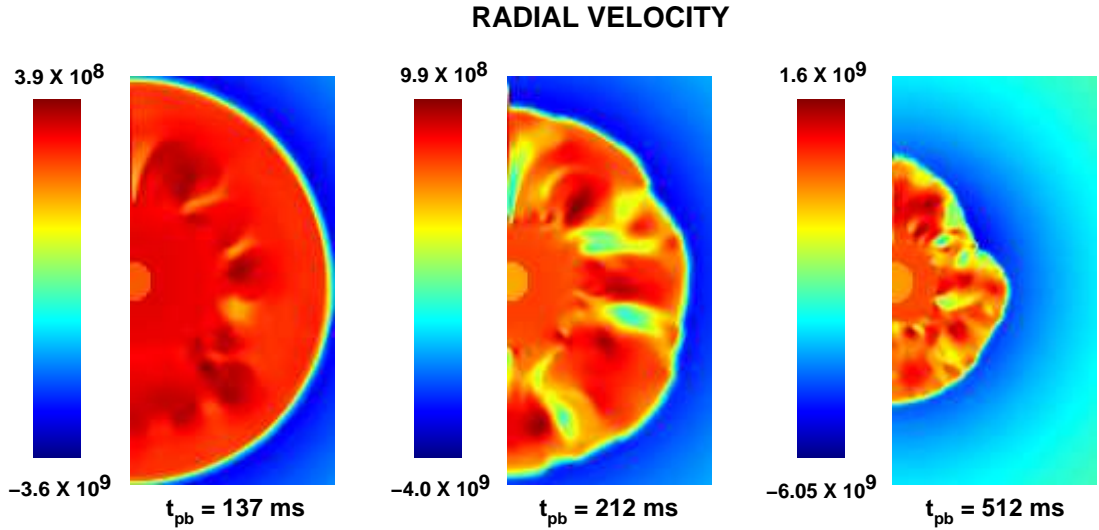,width=5.7in,angle=0}
\caption[Two-dimensional radial velocity]
{\protect\small \protect\baselineskip 10pt \label{nfig5a}
Two-dimensional radial velocity plots showing the
evolution of neutrino-driven convection in our 15 ${\rm M}_{\odot}$
model, at 137, 212, and 512 ms after bounce.
}
\end{figure}

\renewcommand{\thefigure}{\arabic{numbercount}}
\setcounter{numbercount}{13}

\begin{figure}
\centering
\epsfig{file=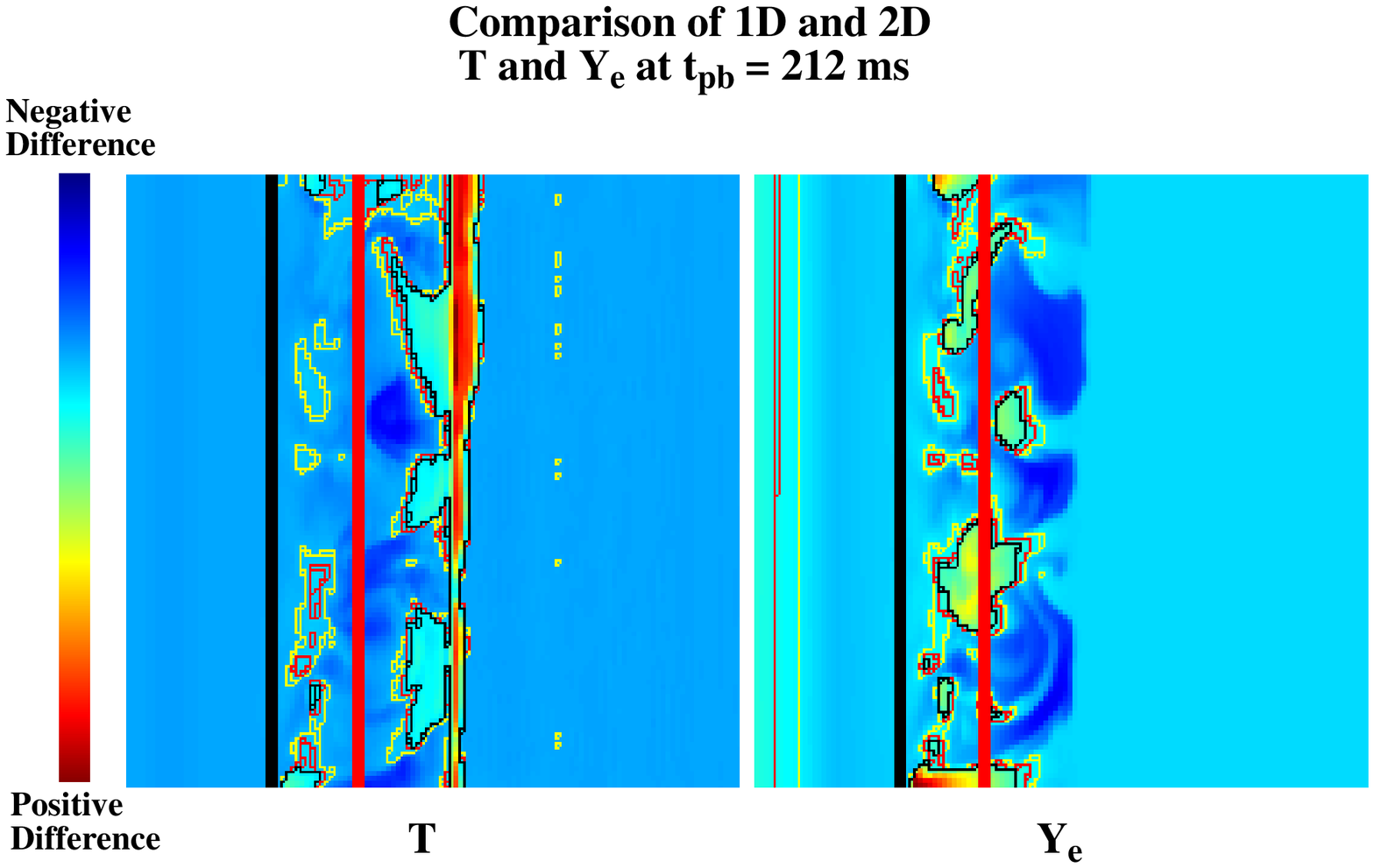,width=5.7in,angle=0}
\caption[Relative differences]
{\protect\small \protect\baselineskip 10pt \label{nfig13}
Plots showing the fractional differences between our
one- and two-dimensional {\small PPM} simulations.
Shown are differences in temperature and electron fraction.
The panels are a Cartesian mapping of our two-dimensional
grid, with 128 angles along the $y$-axis and 128 radii along the
$x$-axis. The contours mark fractional differences of 3\%
(yellow), 6\% (red), and 9\% (black). The thick vertical
lines mark the angle-averaged neutrinosphere and gain radii,
in black and red, respectively.
}
\end{figure}
\begin{figure}[t]
\vspace{3.0in}
\end{figure}
\clearpage